\documentclass{elsart}
\usepackage{natbib,epsfig,fancybox,latexsym}
\usepackage{amsfonts,amssymb}
\usepackage{hhline,eufrak,euscript,delarray}
 
\newcommand{\ENDPROOF}{\hfill \qed}
\def\coeff#1{\left[ #1 \right]}
\def\V{\vartheta}
\def\rbar{\relbar\!\!\relbar} 

\def\gr#1{\widehat{#1}} 
\def\mg#1{#1}
\def\sth#1{\underline{#1}}

\def\PW#1{$( {\mathcal{P}}_{#1,\gr{W}} )$} 
\def\PS#1{$( {\mathcal{P}}_{#1,\gr{S}} )$}
\def\PJ#1{$( {\mathcal{P}}_{#1,\gr{J}} )$}

\def\aut#1{#1}
\def\rond{$\bullet$}
 
\def\cpt#1{ c_{#1}^{(C_3)} }
\def\cpxi#1{ c_{#1}^{(\xi)} } 

\def\betaxi#1{\beta_{#1}^{(\xi)}}
\def\Ckt#1{{\mathcal{C}}_{#1}^{(C_3)}}
\def\Ckxi#1{{\mathcal{C}}_{#1}^{(\xi)}}
\def\Lambdat#1{{\Lambda}_{#1}^{(C_3)}}
\def\Lambdaxi#1{{\Lambda}_{#1}^{(\xi)}}
\def\Omegat#1{{\Omega}_{#1}^{(C_3)}}
\def\Omegaxi#1{{\Omega}_{#1}^{(\xi)}}
\def\xx{\mathbb{X}}
\def\Rhoxi#1{{\rho}_{#1}^{(\xi)}}
\def\Zt#1{ Z_{#1}^{(C_3)} }
\def\Yt#1{ Y_{#1}^{(C_3)} }

\def\EPSILON\epsilon
\def\VAREPSILON\varepsilon

\onecolumn
\begin{document}
\runauthor{Ravelomanana, Thimonier}
\begin{frontmatter}
\title{Forbidden Subgraphs in Connected Graphs\thanksref{X}}
\author[Paris-Nord]{Vlady Ravelomanana}
\author[Amiens]{Lo\"{y}s Thimonier}
\thanks[X]{First version of this paper appeared in the 4-th Latin American
  Theoretical INformatics Conference -- Punta del Este -- Uruguay,
 April 2000 and some parts of this paper appeared in the
13-th International Conference on Formal Power Series and Algebraic
Combinatorics -- Arizona -- USA, May 2001 (cf. \cite{RT00a,RT01a}). This 
research was done while the first author was at
  LaRIA, Amiens -- France.} 

\thanks[EMAIL]{Corresponding authors. Email:
  vlad@lipn.univ-paris13.fr, thimon@laria.u-picardie.fr .}

\address[Paris-Nord]{LIPN UMR 7030, Universit\'e de Paris-Nord\\
99, Avenue J.~B. Cl\'ement. F 93430 Villetaneuse France}

\address[Amiens]{LaRIA EA 2083, Universit\'e de Picardie \\
5, Rue du Moulin-Neuf. 80000 Amiens France}

\date{February 2002}
\maketitle

\begin{abstract}
\noindent Given a set $\xi=\{H_1,H_2,\cdots\}$ of connected non acyclic 
graphs, a $\xi$-free graph is one which does not contain any member of $%
\xi$ as copy. Define the excess of a graph as the difference between
its number of edges and its number of vertices. 
Let ${\gr{W}}_{k,\xi}$ be theexponential generating function (EGF for brief) 
of connected $\xi$-free graphs of excess
equal to $k$ ($k \geq 1$). For each fixed $\xi$, 
a fundamental differential recurrence
satisfied by the EGFs ${\gr{W}}_{k,\xi}$ is derived. 
We give methods on how to
solve this nonlinear recurrence for the first few values of $k$ by means of
graph surgery. We also show that for any finite collection $\xi$ of
non-acyclic graphs, the EGFs ${\gr{W}}_{k,\xi}$ are 
always rational functions of the generating function, $T$, 
of Cayley's rooted (non-planar) labelled
trees. From this, we prove that almost all connected graphs with $n$ nodes
and $n+k$ edges are $\xi$-free, whenever $k=o(n^{1/3})$ and $|\xi| < \infty$
by means of Wright's inequalities and saddle point method. Limiting
distributions are derived for sparse connected $\xi$-free components
that are present when a random graph on $n$ nodes has approximately
$\frac{n}{2}$ edges. In particular, the probability distribution
that it consists of trees, unicyclic components, $\cdots$, $(q+1)$-cyclic
components all $\xi$-free is derived. Similar results are also obtained
for multigraphs, which are graphs where self-loops and
multiple-edges are allowed.
\end{abstract}


\end{frontmatter}
\typeout{SET RUN AUTHOR to \@runauthor}

\section{Introduction}

We consider here labelled graphs, i.e., graphs with labelled
vertices, undirected edges and without self-loops or multiple edges as well
as labelled \textit{multigraphs} which are labelled graphs with self-loops 
and/or multiple edges. A $(n,q)$ graph (resp. multigraph) is one having $n$ 
vertices and $q$ edges.

On one hand, classical papers \cite{ER59, ER60, FKP89, JKLP93} provide
algorithms and analysis of algorithms that deal with random graphs or
multigraphs generation, estimating relevant characteristics of their
evolution. Starting with an initially empty graph of $n$ vertices, we enrich
it by successively adding edges. As random graph evolves, it displays a
phase transition similar to the typical phenomena observed with percolation
process. On the other hand, various authors such as \aut{Wright}
\cite{Wr77, Wr80} or \aut{Bender, Canfield} and \aut{McKay} 
\cite{BCM90, BCM92} studied exact enumeration or 
asymptotic properties of labelled
\textit{connected} graphs.

A lot of research is devoted to graphs not containing
a prefixed set of subgraphs as copies and various approaches
exist for these problems. Most of them, following \aut{Erd\"{o}s}
and R\'enyi's seminal papers \cite{ER59, ER60}, are
 probabilistic; moment methods, tail inequalities, or
probabilistic inequalities are then essential as well explained
in \cite{Bollobas}. These approaches take
advantage over enumerative ones by allowing treatments
under the edges independence assumption \cite{Bollobas}.
The situation changes radically if we consider connected components,
and results relative to connectedness are few. Related
works include \cite{Wr77, Wr78, Wr80, BCM90, BCM92, BCM97, FKP89, JKLP93}

Let $H$ be a fixed connected graph;
by a \textit{copy of $H$}, we mean any subgraph, 
not necessarily induced, isomorphic to $H$. 
Let $\mathcal{F}$ be a family of graphs none of which contains a copy
of $H$. In this case, we say that the family $\mathcal{F}$ 
is \textit{$H$-free}. Otherwise, a graph 
containing a copy of $H$ is called a \textit{supergraph}
of $H$. The highly non-trivial task of enumerating 
\textit{triangle-free} or \textit{quadrilateral-free} 
components goes back to the book of
\aut{Harary} and \aut{Palmer} \cite{HP73}.

Mostly forbidden configurations are triangle, quadrilateral, ...,
$C_{p}$, $K_{p}$, $K_{p,q}$ or 
any combination of them 
(see \cite[Chapter IV]{Bollobas}, \cite[Chapter III]{JLR00}). 
$C_{p}$ shall always denote the
cycle on $p$ vertices, $K_{p}$ the complete graph with $p$ vertices and $%
K_{p,q}$ the complete bipartite graph with $p$ vertices on the first side
and $q$ vertices on the second side. For example, we can work with the
family of graphs which do not contain a copy of   
triangle ($C_{3}$) or of $K_{3,3}$, i.e., 
$\{C_{3},K_{3,3}\}$-free graphs. Following the authors of \cite{FKP89},
we refer as \textit{bicyclic} graphs all connected
graphs with $n$ vertices and $(n+1)$ edges 
and in general $(q+1)$-\textit{cyclic} graphs are
connected $(n,n+q)$ graphs. If we define the \textit{excess} of a graph as
the difference between its number of edges and its number of vertices,
\textit{$(q+1)$-cyclic} graphs are referred also as
\textit{$q$-excess} connected graphs. 
In general, we refer as \textit{multicyclic} a
connected graph which is not acyclic. The same nomenclature holds for
multigraphs. More generally, denote by $\xi
=\{H_{1},H_{2},H_{3},...\}$ a set of connected multicyclic graphs (resp.
multigraphs); a ${\xi}$\textit{-free} graph is then one which 
does not contain any copy of $H_i$ for all $H_i \in \xi$ as subgraph.
Throughout this paper, unless explicitly mentioned, $\xi$ denotes
a \textit{finite} set of forbidden configurations.
 
Our aim in this paper is 
\begin{itemize}
\item[1.]  to study randomly generated  graphs
with $n$ vertex  and approximately $\frac{n}{2}$ edges 
focusing our attention
on  the appearance or not of the forbidden configurations,
\item[2.]  to compute the asymptotic number of $\xi$-free connected
  graphs when $\xi$ is finite. 
\end{itemize}
The results obtained here show that some characteristics of random
generation as well as asymptotic enumeration of labelled
graphs or multigraphs,  can be read within the forms of the
exponential generating functions (EGF for short) of the sparse components.
In fact, denote by  $\gr{W}_{k}$ ($k \geq -1$) the EGFs 
of $(k+1)$-cyclic (connected) graphs.
In a series of important papers, \cite{Wr77, Wr78, Wr80}, 
E.~M. Wright proved that $\gr{W}_{k}(z)$ $(k \geq 1)$, where $z$
is the variable marking the number of vertices in the graph, 
can be expressed as finite sums of power of
$1/(1-T(z))$ where $T(z) = \sum_{n \geq 1} n^{n-1} \frac{z^n}{n!}$
is the EGF for rooted labelled trees \cite{Cay89, Moo67}.
Starting with a functional equation satisfied 
by our $(k+1)$-cyclic
$\xi$-free graphs; we will show that their EGF, denoted $\gr{W}_{k,\xi}$,
have the same global forms as 
those of  $(k+1)$-cyclic graphs, i.e.,
$\gr{W}_{k}$. These forms will allow us to study random
graphs without forbidden configurations and also to
enumerate asymptotically connected components of these objects under
some restrictions. Similar results related to multigraphs will be 
treated and carried along this paper, in parallel. Since our results
 concern  graphs and multigraphs,
we will be frequently assuming throughout this paper that the
term \textit{component} is the general term for  connected  graph
as well as for connected multigraph.

\subsection{Asymptotic number of $\xi$-free $(n,n+o(n^{1/3}))$ components}
In the first part of this paper, we will compute 
the asymptotic number of triangle-free connected $(n,n+k)$-graphs, whenever
$k=o(n^{1/3})$.
To do this, we will rely heavily on the results in \cite{Wr80} to prove that
the power series $\gr{W}_{k,C_3}$ satisfy the same
inequalities as for $\gr{W}_k$ which we shall call here
 ``\textit{Wright's inequalities}''. Next, 
we will investigate the asymptotic
 behavior of the coefficient of $z^n$ in $\frac{1}{(1-T(z))^{k(n)}}$
(where $T$ is the EGF for Cayley's rooted labelled trees)
 by means of saddle point method. The combination of these computations
will permit us to show 
\textit{almost all} 
connected $(n,n+o(n^{1/3}))$ graphs,
i.e., connected graphs with $n$ vertices and $n+o(n^{1/3})$ edges are
triangle-free. These asymptotic results are related
 to the interesting problems posed by \aut{Harary} and \aut{Palmer}
in their reference book (see \cite[Sect. 10.4, 10.5 and 10.6]{HP73}). 
The purpose of this part is also to introduce methods by which the asymptotic 
number of connected $\xi$-free $(n,n+k)$  graphs 
can be computed systematically, whenever $k=o(n^{1/3})$.

\subsection{Forbidden subgraphs in random $(n,\frac{n}{2})$ components}
The two models of graph evolution, explicitly introduced 
in \cite{FKP89}, are considered in the second part of this note,  in
order to generate randomly graphs and multigraphs. We will study the
structure of evolving graphs and multigraphs when edges are added one at
time and at random, mainly looking at the presence or absence of certain
configurations. In \cite[Theorem 5]{JKLP93}, the authors proved that
the probability that a random graph or multigraph with $n$ vertices and $%
\frac{n}{2}+O(n^{1/3})$ edges has $r_{1}$ bicyclic components, $r_{2}$
tricyclic components, ..., $r_{q}$ $(q+1)$-cyclic components 
and no components of higher-cyclic order is 
\begin{equation}
\left( \frac{4}{3}\right) ^{r} %
\sqrt{\frac{2}{3}}\frac{b_{1}^{r_{1}}}{r_{1}!}%
\frac{b_{2}^{r_{2}}}{r_{2}!} \cdots %
\frac{b_{q}^{r_{q}}}{r_{q}!}\frac{r!}{%
(2r)!}+O(n^{-\frac{1}{3}})  \label{JKLP93Theorem5}
\end{equation}
where $r=r_{1}+2r_{2}+ \cdots + qr_{q}$ and 
the $b_i$ are \textit{Wright's constants} also found
by \aut{Louchard} and \aut{Tak\'acs}
($b_{1}=\frac{5}{24}, \, b_{2}=\frac{5}{16}$, \, ...), and
are involved in an important series of papers
 \cite{Lo84a,Lo84b,Vo87,Ta91a,Ta91b,JKLP93,Sp97,FPV98}.

Given a finite collection 
$\xi =\{H_{1},H_{2},H_{3},...,H_{q}\}$ of
multicyclic connected components, with slight 
modifications of the results in 
\cite{JKLP93}, we show that for a random graph or multigraph with $n$ 
vertices and
$m(n)=\frac{n}{2}(1+\mu n^{-\frac{1}{3}})$ edges,
 $|\mu |\leq n^{1/12}$
(in this paper, we will often choose $\mu=O(n^{- \frac{1}{3} } ) $ so
$m(n)= \frac{n}{2} +O(n^{\frac{1}{3}})$), the
probability of finding only acyclic and unicyclic components without
copy of $H_{i}$, $\forall H_{i}\in \xi $, is
asymptotically the same value as for ``general'' 
random graphs \textit{times}
${\exp{\left( -\sum_{k\in \Theta }\frac{1}{2p}\right)} }$ where $%
\Theta $ is the subset (possibly empty) of the lengths of all 
polygons in $\xi$:
 $\Theta=\{p,H_{i}\in \xi $ and $H_{i}$ is a $p$-\textit{gon}$\}$. 
For example, if $%
\xi =\{C_{3},C_{4}\}$, $\Theta=\{3,4\}$ and
 the probability that a random graph or a multigraph
with $n$ vertices and $\frac{n}{2}+O(n^{1/3})$ edges has only trees and
unicyclic components without \textit{triangles} 
or \textit{quadrilaterals} as
induced subgraphs is
\begin{equation}
\sqrt{\frac{2}{3}}e^{-\frac{1}{6}-\frac{1}{8}}  \sim 0.6099\cdots \,  .
\label{SIMPLEWITHOUTC3C4}
\end{equation}
Recall that an \textit{elementary contraction} of
 a graph $G$ is obtained by 
identifying two adjacent points $x$ and $y$, that is, by the removal of $x$
and $y$ and the addition of a new point $z$ 
adjacent to those points to which
$x$ or $y$ were adjacent. Then a graph $G_{1}$ is 
\textit{contractible }to a
graph $G_{2} $ if $G_{2}$ can be obtained from $G_{1}$ by a sequence of
elementary contractions. We show that
 a sufficient condition to change the coefficient $b_{i}$, 
for any $i>0$, of (\ref{JKLP93Theorem5}) in
this probability is to force  $\xi$ to
contains the entire family of graphs \textit{%
contractible} to certain graphs $H_{1}, H_{2}, \cdots$ (in
this case $\xi $ is \textit{infinite}). We then give the corresponding
probability.

\begin{figure}
    \begin{center} \epsfig{file= 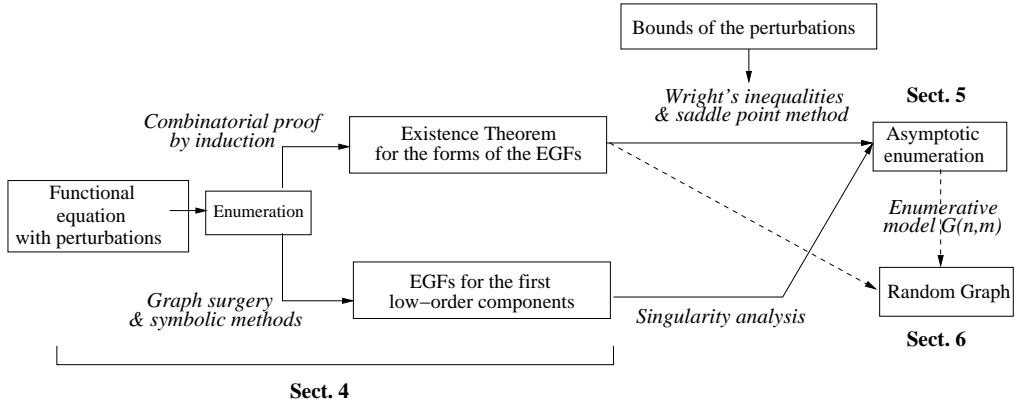,scale=0.50}
    \end{center}
    \caption[sections 4, 5, 6.]
    {Summarizing sections 4, 5, 6 and the methods therein.}
\label{FIG:ABSTRACT456} 
\end{figure}

The ideas of sections 4, 5 and 6 may be summarized by the figure
\ref{FIG:ABSTRACT456}. \\

\subsection{An outline of the paper}
The rest of this paper is organized as follows. In
section 2, we recall some useful definitions and notations
 of the stuff we will encounter
along this document. In section 3, we will work with the
example of bicyclic graphs. The enumeration of these
graphs was discovered, as far as we know, independently by \aut{Bagaev} 
\cite{Bag73} and by \aut{Wright} \cite{Wr77}. The purpose of this example
is two-fold. First, it brings a simple new combinatorial point of view to
the relationship between the generating functions of some \textit{integer
partitions}, on one hand, and \textit{graphs}, on
the other hand. Next, this example gives us ideas, regarding the
 \textit{simplest complex components}, i.e., simplest non-acyclic
components, of what will happen if we force our
 graphs to contain some specific configurations 
(especially the form of the generating functions). In section 4,
we start giving the functional equation satisfied by our 
$\xi$-free connected graphs involving also the first 
components containing copies of
forbidden configurations. This equation is difficult to solve but
leads to the general forms of the EGFs of all $(k+1)$-cyclic $\xi$-free
components. In fact, general combinatorial techniques  are
presented and used to enumerate the first low-order cyclic triangle-free
components. 
 Section 5 presents methods to estimate asymptotically  the number of
 connected components built 
 with  $n$ vertices and $n+k$ edges as $n \rightarrow \infty$ and $k
 \rightarrow  \infty$  but $k=o(n^{1/3})$. The obtained results show that
 \textit{almost all} $(n,n+o(n^{1/3}))$ 
connected components are triangle-free and the methods used
show that this fact can be generalized to any finite set $\xi$
of forbidden subgraphs. 
We then turn on the computation of the
 probability of random graphs/multigraphs without forbidden 
configurations in section 6.
 Along this paper, \textit{triangle-free} graphs
 will be treated as significant example but many results stand for any
\textit{finite} set $\xi$ of forbidden multicyclic graphs or multigraphs.

\section{Notations}

Definitions and tools are given in this section. Because they are mostly
well known, they are quickly sketched. Powerful tools in all
combinatorial approaches, \textit{generating functions} will be used for our
concern. If $F(z)$ is a power series, we write $\left[ z^{n}\right] F(z)$
for the coefficient of $z^{n}$ in $F(z)$. 
We say that $F(z)$ is the \textit{exponential generating function}
 (EGF for brief) for a collection
$\mathcal{F}$ of \textit{labelled} objects 
if $n!\left[ z^{n}\right] F(z)$ is
the number of ways to attach objects in $\mathcal{F}$ that 
have $n$ elements (see for instance \cite{FZvC94} or \cite{Wi90}).

The generating functions for labelled unrooted and labelled rooted 
trees are nice examples of EGFs. The
mathematical theory of labelled trees, as first discussed by \aut{Cayley} 
in 1889 \cite{Cay89} was concerned in their enumeration aspect. This study
initiated the enumeration of labelled graphs. In fact, a labelled tree is a
connected graph with $n$ vertices labelled from $1$ to $n$ and $n-1$ edges.
It is well known that the number of such structures upon $n$ points is $%
n^{n-2}$. Let $T$ be the EGF for labelled rooted trees.
A tree consists of a root to which is attached a set of
rooted subtrees, thus
\begin{equation}
T(z)=z\left( \sum_{n\geq 0}\frac{T(z)^{n}}{n!}\right) = %
\sum_{n\geq 1}n^{n-1}%
\frac{z^{n}}{n!} \, .
\label{EGF-ROOTED-TREES}
\end{equation}
In (\ref{EGF-ROOTED-TREES}), the exponent of the variable $z$ reflects
the number of nodes. One can use 
\textit{bivariate exponential generating function} to count
labelled rooted trees. Throughout this paper, the variable $z$ 
 is the variable recording the number of nodes and 
$w$ is the variable for the number of edges.
For e.g., a tree with $n$ vertices is a connected graph with 
$n-1$ edges and we have
\begin{equation}
T(w,z)=z\, \exp{\left(w \,T(w,z)\right)} = %
\sum_{n>0}(wn)^{n-1}\frac{z^{n}}{n!} \, .
\label{BGF-ROOTED-TREES}
\end{equation}
This bivariate EGF satisfies
\begin{equation}
T(w,z)=\frac{T(wz)}{w} \, .
\label{BGF-EGF-ROOTED-TREES}
\end{equation}
We will denote by $W_{k}$, resp. $\gr{W}_{k}$, the EGF for labelled
multicyclic connected multigraphs, resp. graphs, with $k$ edges more than
vertices. For $k\geq 1$, these EGFs have been computed in \cite{Wr77} 
and in \cite{JKLP93}.
A connected graph is of excess $k$ which is always greater than or
equal to $-1$. Let $\gr{W}_{-1}$ be the EGF of unrooted labelled trees.
One can obtain at generating function level the relation
\begin{equation}
\gr{W}_{-1}(z) = \int_{0}^{z} T(x) \frac{dx}{x}\, ,
\label{ROOTED-UNROOTED}
\end{equation}
which reflects the fact that any
node of an unrooted tree can be taken as the root. The integration
of (\ref{ROOTED-UNROOTED}) leads to the classical relation
\begin{equation}
\gr{W}_{-1}(z) = T(z) - \frac{T(z)^2}{2} \, .
\label{UNROOTED}
\end{equation}
It is convenient to work with bivariate EGFs and 
the bivariate EGFs that enumerate the family
$\gr{\mathcal{W}}_{k}$ of labelled  $k$-excess graphs, for all
$k \geq -1$, can be expressed
using the corresponding univariate EGFs as follows
\begin{equation}
\gr{W}_{k}(w,z) = w^k \gr{W}_{k}(wz) \, . 
\label{UNIVARIATE-BIVARIATE}
\end{equation}
The factor $w^k$ in the right side of (\ref{UNIVARIATE-BIVARIATE}) 
reflects the excess of the component, that is
its number of edges minus its number of vertices. 
The same remark holds between the univariate and bivariate EGFs,
 $\mg{W}_{k}$,  of $k$-excess multigraphs.

Without ambiguity, one can also associate a given configuration of labelled
graph or multigraph with its EGF. For instance, a triangle
 can be labelled in only one way and we have the following
informal relation
\begin{equation}
C_{3} \rightarrow C_{3}(w,z)=\frac{1}{3!}w^{3}z^{3} \, .
\label{TRIANGLE-TRIANGLE-EGF}
\end{equation}
For any given multicyclic component $H$,
 denote by $W_{k,H}$ (resp. $\gr{W}_{k,H}$) the EGF of
multicyclic $H$-free multigraphs (resp. graphs) 
with $k$ edges more than vertices.
In these notations, the second index refers to the forbidden
configuration(s). Recall that a \textit{smooth}
 graph or multigraph is one with all
vertices of degree $\geq 2$ (see \cite{Wr78}). Throughout the rest of this
paper, the ``\textit{widehat}'' notation will be used for EGF of graphs and
``\textit{underline}'' notation corresponds to the 
 \textit{smoothness} of the species. E.g., 
$\underline{\gr{W}_k}$, resp. $\underline{W_k}$,
are EGF for connected $(n,n+k)$ smooth graphs, resp. smooth
multigraphs. 

\begin{rem} \label{REM_MULTIGRAPH}
We follow the authors of  \cite{JKLP93} and the widehat notation will
be used for graphs generating functions. Although, our main concern is graphs,
one can extend the results presented in this paper to multigraphs. In fact,
in the giant paper \cite{JKLP93}, the uniform model of random graphs which
allows self-loops and multiple edges is treated and shown to be easier to
analyze than the classical model of random graphs due to  Erd\"{o}s and R\'{e}nyi
\cite{ER60} since the multigraphs EGFs have better expressions.
\end{rem}

We need additional definitions corresponding to the first appearance
of the forbidden configurations in some random evolving graphs/multigraphs.
For sake of simplicity, we suppose temporarily that $\xi=\{C_3\}$.
Consider the random graph process which starts 
with $n$ initially disconnected
nodes. When enriching it by  successively adding edges,
 one at time and at random, the first time a new copy of
 triangle is created with the last added edge in some 
connected component, there are two possibilities: 
\begin{itemize}
\item[1.] the last edge creates \textit{exactly}
 one and only one triangle,
\item[2.] there are many occurrences of triangles but sharing
the last added edge which deletion will 
suppress all copies of triangle in
the considered component. We shall call this sort of configuration
``\textit{juxtaposition}'' of triangles.
\end{itemize}
The same nomenclature holds when considering a set $\xi$ of forbidden
configurations. For example if $\xi=\{C_3,C_4\}$, a ``\textit{house}''
can appear in some component. More formally, we have the following
reformulation related to these kinds of construction:
\begin{defn} \label{DEF_JUXTA}
Given a subset $\{H_{i_1}, \, H_{i_2}, \, \cdots, \, H_{i_q}\}$
of $\xi$, we define the \textit{juxtaposition} of 
$H_{i_1}, \, H_{i_2}, \, \cdots, \, H_{i_q}$
 as a subgraph containing at least one
copy of each $H_{i_j}$ but such that there exists an edge
which deletion will suppress all the occurrences of 
$H_{i_1}, \, H_{i_2}, \, \cdots, \, H_{i_q}$. When there exists
$s$ shared edges such that the deletion of any of them will suppress
all the occurrences of $H_{i_1}, \, H_{i_2}, \, \cdots, \, H_{i_q}$,
we define this specific configuration as a $s$-juxtaposition.
\end{defn}

\begin{exmp}
We have the figure \ref{HOUSE}
 depicting a $1$-juxtaposition of $C_3$ and $C_4$,
representing a ``house''. In figure \ref{2K4}, we have
 a $1$-juxtaposition and a $3$-juxtaposition 
of two $K_4$.
\begin{figure}[h]
  \begin{minipage}[t]{6.0cm}
    \begin{center} \epsfig{file=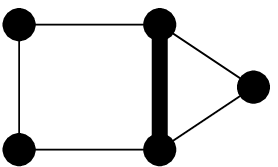,scale=0.5}
    \end{center}
    \caption[$1$-juxtaposition of $C_3$ and $C_4$ ($\xi=\{C_3, \, C_4\}$)]
    {The ``house'': $1$-juxtaposition of $C_3$ 
 and $C_4$ ($\xi=\{C_3, \, C_4\}$).}
        \label{HOUSE}
  \end{minipage}
\hfill
  \begin{minipage}[t]{7.0cm}
    \begin{center} \epsfig{file=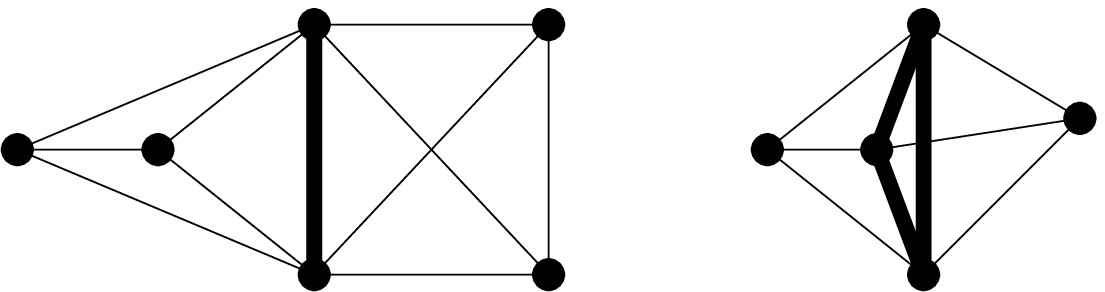,scale=0.5}
    \end{center}
    \caption[$1$-juxtaposition and $3$-juxtaposition of $K_4$]
    {$1$-juxtaposition and $3$-juxtaposition of $K_4$ ($\xi=\{K_4\}$).}
        \label{2K4}
  \end{minipage}
\end{figure}
\end{exmp}
\begin{defn} \label{DEF:JKXI}
For any $H \in \xi$, denote by $\gr{S}_{k,H}$ the
EGF of $(k+1)$-cyclic graphs with exactly one copy of $H$
(copies of other graphs of $\xi$ are not allowed).
Define  by $\gr{S}_{k,\xi} = \sum_{H \in \xi} \gr{S}_{k,H}$,
the EGF of $(k+1)$-cyclic graphs with one occurrence of a member of $\xi$.
For any subset $\xi^{'} \subseteq \xi$, denote by $\gr{J}_{k,\xi^{'}}^{(p)}$
the EGF of $p$-juxtaposition of $\xi^{'}$. We let
$\gr{J}_{k,\xi} = \sum_{\xi^{'} \subseteq \xi} \sum_{p} %
p \, \gr{J}_{k,\xi^{'}}^{(p)}$.
Respectively, $S_{k,\xi}$ and $J_{k,\xi}$ are the
EGFs for multigraphs with the same characteristics.
\end{defn}

%
%
%

Furthermore, denote by $\V_w$, resp. $\V_z$,
the differential operator $w \frac{\partial}{\partial w}$, resp. $z
\frac{\partial}{\partial z}$. The operator $\V_w$
corresponds to marking an edge of a graph (or a multigraph).
Similarly, $\V_z$ corresponds to marking a vertex .
For the use of pointing and marking, we refer to 
\cite{GJ83} and  for general techniques 
concerning graphical enumerations we refer to \cite{HP73}.

The following observation will take its importance as we will
see later:
\begin{rem} \label{JKXI}
$\gr{J}_{k,\xi}$ is the EGF of $(k+1)$-cyclic graphs
with a shared edge of the juxtaposition marked.
\end{rem}

\begin{rem} \label{PRECEDENCE} Throughout this paper, we will 
frequently use the following notation when comparing the
coefficients of two generating functions.
If $A$ and $B$ are two formal power series such that
for all $n \geq 0$ we have $\coeff{z^n} A(z) \leq \coeff{z^n}B(z)$ then
we denote this relation $A \preceq B$ (or $A(z) \preceq B(z)$).
\end{rem}

\section{The link between the EGF of bicyclic graphs and integer partitions}
\label{SEC:LINK}

At least in 1967, there were $10$ different proofs for the EGF for trees
according to the paper of \aut{Moon} \cite{Moo67} and $16$ proofs
related in \cite{K73}. Then, 
\aut{R\'{e}nyi} 
\cite{Ren59} found the formula to enumerate unicyclic graphs which 
can be expressed in terms of the generating function of rooted 
labelled trees,  namely
\begin{equation}
\gr{W}_{0}(z)=\frac{1}{2}\ln{\frac{1}{1-T(z)}}%
-\frac{T(z)}{2}-\frac{T(z)^{2}}{4} \, .
\label{UNICYCLIC-GRAPH}
\end{equation}
We refer here to the symbolic methods developed in \cite{FS96}
for modern computation of formulae like (\ref{UNICYCLIC-GRAPH}).
The formula for unicyclic multigraphs is very similar
and there are terms due to self-loops and multiple edges
\begin{equation}
W_0(z)=\frac{1}{2}\ln{\frac{1}{1-T(z)}} \, .
\label{UNICYCLIC-MULTIGRAPH}
\end{equation}
It may be noted that in some connected graphs, as well as
multigraphs the number of edges exceeding the number of vertices 
can be seen as useful enumerating parameter. 
The term \textit{bicyclic} graphs, appeared
first in the seminal paper of \aut{Flajolet} \textit{et al.} 
\cite{FKP89} followed
few years later by the huge one of \aut{Janson} \textit{et al.} 
\cite{JKLP93} and
was concerned with all connected graphs with $(n+1)$ 
edges and $n$ vertices.
The authors of these documents choose then the word \textit{bicyclic}
 for connected component
 which is constructed by adding a random
edge to a unicyclic component.
 \aut{Bagaev} \cite{Bag73} first found a method to count
such graphs. His method of \textit{shrinking-and-expanding} graphs is
well explained in \cite{BV98}. \aut{Wright} \cite{Wr77} found a recurrent
formula well adapted for formal calculation to compute the number of all
connected graphs of excess $k$ (for all $k \geq 1$). Our aim 
in this section is to show that the
problem of the enumeration of  \textit{bicyclic graphs} can also be solved
with techniques involving integer partitions. We present here
a simple treatment very close to the Wright's method 
 as a warm-up for the forthcoming results in the next sections.
   
\begin{figure}[h]
  \begin{minipage}[t]{7.0cm} 
    \begin{center} \epsfig{file=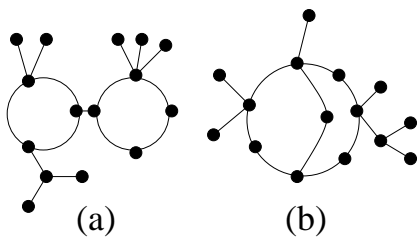,scale=0.5}
    \end{center}
    \caption[Examples of bicyclic components.]
    {Examples of bicyclic components.}
    \label{examples_bicyclic}
  \end{minipage}
  \begin{minipage}[t]{7.0cm}
    \begin{center} \epsfig{file=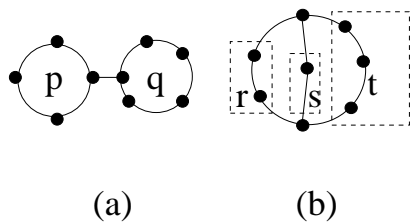,scale=0.5}
    \end{center}
    \caption[Smooth bicyclic components.]
    {Smooth bicyclic components.}
    \label{smooth_bicyclic}
  \end{minipage}
\end{figure}

Given a fixed set of $n$ vertices, there exist two types 
of graphs which are connected and have $(n+1)$ edges as 
described in the figure \ref{examples_bicyclic}. \\

\begin{figure}[h]
  \begin{center} \epsfig{file=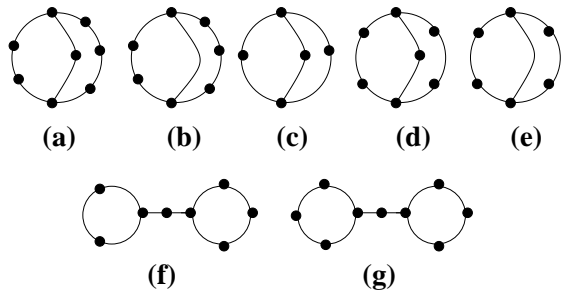,scale=0.40}
  \end{center}
  \caption[The different basic smooth bicyclic graphs.]
  {The different basic smooth bicyclic graphs.}
  \label{basic_bicyclic}
\end{figure}

\noindent
\aut{Wright} \cite{Wr77} showed with his \textit{reduction} method that
the EGF of all multicyclic graphs, namely bicyclic graphs, can be expressed
in terms of the EGF of labelled rooted trees. In order to 
count the number of
ways to label a graph, we can repeatedly \textit{prune }it by suppressing
recursively any vertex  of degree $1$. We then remove as many vertices as
edges. As these structures present many symmetries, our experiences suggest
us so far that we ought to look at our previously described object without
symmetry and without the possible rooted subtrees.
There are 
\[
{n \choose p}{ {n-p} \choose q}%
 \frac{(p-1)!}{2}p\frac{(q-1)!}{2}q(n-p-q)!=\frac{n!}{4} 
\]
manners to label the graph represented by the figure 
\ref{smooth_bicyclic} (a)
whenever $p\neq q$. In the graph of figure \ref{smooth_bicyclic} (b),
if $r \neq s$, $s \neq t$, $t \neq r$, there are $\frac{n!}{2}$ 
ways to label the graph. Note that these results are
independent from the size of the subcycles. One can obtain all smooth
bicyclic graphs after considering possible symmetry criterions. In figure
\ref{smooth_bicyclic} (a), if the subcycles have the same length, $p=q$, a
factor $\half$ must be considered and we have $n!/8$ ways to label the
graph. Similarly, the graph of figure \ref{smooth_bicyclic} (b) 
can have the 3 arcs with the same number of vertices. 
In this case, a factor $1/6$ is introduced. If only two arcs have the same
number of vertices, we need a symmetrical factor $1/2$. Thus, the
enumeration of smooth bicyclic graphs can be viewed as specific problem of
integer partitioning into 2 or 3 parts following the dictates of the basic
graphs in figure \ref{basic_bicyclic}. \\

\noindent
With the same notations as in \cite{Co70}, denote by $P_{i}(z)$,
respectively $Q_{i}(z)$, the generating functions of the number of
partitions of an integer in $i$ parts, respectively in $i$ different parts.
Let $\underline{\gr{W}_{1} }(z)$ be the univariate EGF for smooth bicyclic graphs, then we
have $\underline{\gr{W}_{1}}(z)=f\big(P_2(z),P_3(z),Q_2(z),Q_3(z)\big)$, i.e.,
\begin{equation}
\begin{array}{cc}
\underline{\gr{W}_{1}}(z)= & %
\underbrace{\frac{1}{2}z^{2}(Q_{3}(z)+Q_{2}(z))}_{%
\mbox{figures \ref{basic_bicyclic} (a), \ref{basic_bicyclic} (b)}} %
+ \underbrace{\frac{1}{12}\frac{z^{5}}{1-z^{3}}}_{%
\mbox{\ref{basic_bicyclic} (c)}} \\ 
& +\underbrace{\frac{1}{4}\left( \frac{z^{4}}{1-z^{2}}+\frac{z^{5}}{%
(1-z)(1-z^{2})}-\frac{z^{5}}{(1-z^{3})}\right) }_{%
\mbox{\ref{basic_bicyclic} (d), \ref{basic_bicyclic} (e)}} \\ 
& +\underbrace{\frac{1}{4}\frac{z^{6}}{(1-z)^{2}(1-z^{2})}}_{%
\mbox{\ref{basic_bicyclic} (f)}}+%
\underbrace{\frac{1}{8}\frac{z^{5}}{(1-z)(1-z^{2})}}_{%
\mbox{\ref{basic_bicyclic} (g)}} \, . 
\end{array}
\label{DECOMPOSITION-PARTITION}
\end{equation}
In formula (\ref{DECOMPOSITION-PARTITION}) or equivalently $%
\underline{\gr{W}_{1}}(z)=\frac{z^{4}}{24}\frac{(6-z)}{(1-z)^{3}}$, the
denominator $\frac{1}{(1-z)^{3}}$ denotes the fact that there is at most $3$
arcs or $3$ \textit{degrees of liberty} of integer partitions of the
vertices in a bicyclic graph. The same remark holds for the denominators $%
\frac{1}{(1-T(z))^{3k}}$ in Wright's formulae \cite{Wr77} for all 
$(k+1)$-cyclic connected labelled graphs. 
To get the whole EGF for bicyclic graphs, we have to substitute
$z$ by $T(z)$ in $\sth{\gr{W}_1}(z)$ in order to replace all
(shrinked) vertices of the smooth graphs by labelled rooted trees.
The form of these EGF 
takes its importance when  studying
the asymptotic behavior of random graphs or multigraphs with a 
given excess. In fact, the known expansion of the Cayley's 
function, $T$, at its singularity $z=\frac{1}{e}$ is 
(see \cite{KP89, FO90, FS+})
\begin{equation}
T(z) = 1 - \sqrt{2}\delta + \frac{2}{3}{\delta}^2 %
- \frac{11}{36}\sqrt{2}{\delta}^3 + \cdots \, , \, \, (\delta=\sqrt{1-ez})\, .
\label{SINGULARITY-CAYLEY}
\end{equation}
As the EGFs of multicyclic components can be expressed
in terms of $T$, the key point of
 their characteristics corresponds directly to
 the analytical properties of \textit{tree
polynomial} $t_{n}(y)$ defined as follow
\begin{equation}
\frac{1}{(1-T(z))^y}=\sum_{n\geq 0}t_{n}(y)\frac{z^{n}}{n!} \, .
\label{TREE-POLYNOMIAL}
\end{equation}
($t_{n}(y)$ is a polynomial of degree $n$ in $y$.) \aut{Knuth} 
and \aut{Pittel} \cite{KP89} studied their properties. For
\textit{fixed} $y$ as $n\rightarrow \infty $, we have (see 
\cite[lemma 2]{KP89})
\begin{equation}
t_{n}(y)=\frac{\sqrt{2\pi }n^{(n-1/2+y/2)}}{2^{y/2}\Gamma (y/2)}%
+O(n^{n-1+y/2})  \, .
\label{TREE-POLYNOMIAL-ASYMPT}
\end{equation}
This equation tells us that in the EGF, $\gr{W}_{1}$ of bicyclic graphs,
expressed here as a sum of powers of $1/(1-T(z))$
\begin{eqnarray}
\gr{W}_{1}(z) & = & \frac{T(z)^{4}}{24}\frac{(6-T(z))}{(1-T(z))^{3}} \cr
                   & = & \frac{5}{%
24}\frac{1}{(1-T(z))^{3}}-\frac{19}{24}\frac{1}{(1-T(z))^{2}} + %
\cdots \, , 
\label{ANOTHER-DVPT-BICYCLIC}
\end{eqnarray}
only the coefficient $\frac{5}{24}$ of $t_{n}(3)$ is
asymptotically significant.

\section{Functional equation for $\xi$-free graphs/multigraphs
and the forms of their EGFs}
\subsection{Differential recurrence for $\xi$-free components}
\label{SUBSUB:FUNCTIONAL}
EGFs of triangle-free unicyclic components can be easily obtained
when avoiding cycle of length $3$ in the general formulae for 
unicyclic graphs (\ref{UNICYCLIC-GRAPH}), resp. multigraphs 
(\ref{UNICYCLIC-MULTIGRAPH}). Denote respectively by 
$W_{0,C_{3}}$ and $\gr{W}_{0,C_{3}}$ the EGFs for unicyclic multigraphs and
graphs without triangle ($C_{3}$), we have
\begin{equation}
W_{0,C_{3}}(z)=\frac{1}{2}\ln{\frac{1}{1-T(z)}}-\frac{T(z)^{3}}{6} \, ,
\label{ACYCLIC-MC3FREE}
\end{equation}
\begin{equation}
\gr{W}_{0,C_{3}}(z)=\frac{1}{2}\ln{\frac{1}{1-T(z)}}-\frac{T(z)}{2}-\frac{%
T(z)^{2}}{4}-\frac{T(z)^{3}}{6} \, .
\label{ACYCLIC-GC3FREE}
\end{equation}
Enumerating components of higher 
cyclic order without triangle is much more difficult. However, we 
have the following lemma: 
\begin{lem} \label{LEMMA0} 
For all $i \geq -1$, denote by  $\gr{W}_{i,C_3}$ the EGF for
triangle-free $(i+1)$-cyclic graphs. Let $\gr{S}_{i,C_3}$ and 
 $\gr{J}_{i,C_3}$ be the EGFs described as in definition
\ref{DEF:JKXI}.
Then, the bivariate EGFs $\gr{W}_{k+1,C_3}$, 
$\gr{S}_{k+1,C_3}$, $\gr{J}_{k+1,C_3}$ and
$\gr{W}_{p,C_3}$ for 
$-1 \leq p \leq k$ are related by the differential recurrence:
\begin{eqnarray}
\V_w \gr{W}_{k+1,C_3} \, \, \, &+&  \, \, 3 \gr{S}_{k+1,C_3}
\, \,  + \, \, \gr{J}_{k+1,C_3} \, \, = \, \,  %
 \, \, w \Big( \frac{{\V_z}^2 - \V_z}{2} - \V_w \Big) \gr{W}_{k,C_3} \cr
 & + & \, \, w \left(%
\sum_{ -1\leq p \leq q \leq k+1, \, p+q=k} %
               \frac{1}{1+\delta_{p,q}} (\V_z \gr{W}_{p,C_3})%
                (\V_z \gr{W}_{q,C_3}) \right) \, 
\label{FUNCTIONAL_TRIANGLE_GRAPH}
\end{eqnarray} 
where $\delta_{p,q}=1$ iff $p=q$, otherwise $\delta_{p,q}=0$.
Similarly, we have for multigraphs (with the same parameters): 
\begin{eqnarray}
\V_w {W_{k+1,C_3}} \, \, \, &+&  \, \, 3 {S_{k+1,C_3}}
\, \,  + \, \, {J_{k+1,C_3}} \, \, = \, \,  %
 \, \, w  \Big(\frac{{\V_z}^2}{2} {W_{k,C_3}}\Big) \cr
 & + & \, \, w \left(\sum_{-1\leq p \leq q \leq k+1, \, p+q=k} %
               \frac{1}{1+\delta_{p,q}} (\V_z {W_{p,C_3}})%
                (\V_z {W_{q,C_3}})  \right) \, .
\label{FUNCTIONAL_TRIANGLE_MULTIGRAPH}
\end{eqnarray}
\end{lem} 
\noindent \textbf{Proof.}  There are two ways to obtain a 
$(k+2)$-cyclic component from components of lower cyclic order,
which are in the right part of (\ref{FUNCTIONAL_TRIANGLE_GRAPH}) 
and are assumed  to be triangle-free. For multigraphs,
we have to employ the combinatorial
operation $\frac{{\V_z}^2}{2}$. 

First of all, consider a triangle-free 
$(k+1)$-cyclic component. To add a new edge to this component,
we have to choose two vertices,
different and  already not adjacent for graphs,  
and not necessarily different for multigraphs. For graphs,
the combinatorial operator used to choose two different vertices
is $ \frac{{\V_z}^2 - \V_z}{2} $. Then, we have to avoid the adjacent 
vertices by means of the operator $-  \V_w$ (see  \cite[Section 10]{JKLP93}
or  \cite{GJ83} for the use of marking and pointing).
 If the new $(k+2)$-cyclic component contains a triangle, the triangle
can only occur in the following cases:
\begin{itemize}
\item[1.] The new edge creates exactly a triangle. In this case,
 the last added edge is necessarily one of 
the $3$ edges of the new triangle.
\item[2.] The last edge creates many triangles but necessarily
juxtaposed as defined above (definition \ref{DEF_JUXTA}), 
and in this latter case, the last edge
is necessarily the one which is shared
between all the occurrences of triangle.
\end{itemize}
Thus, the left side of (\ref{FUNCTIONAL_TRIANGLE_GRAPH}), resp. of
(\ref{FUNCTIONAL_TRIANGLE_MULTIGRAPH}), distinguishes the last added edge in
the new $(k+2)$-cyclic component.

Next, a $(k+2)$-cyclic triangle-free component can be 
built when creating an edge between a $(p+1)$-cyclic
and a $(q+1)$-cyclic triangle-free components such that
$p+q=k$ and $ -1 \leq p \leq q \leq k+1$ (note that
the case $p=-1$ and $q=k+1$ corresponds to the case
where a tree is attached to a $(k+1)$-cyclic
triangle-free component).
This construction is done by choosing one vertex
belonging to the $(p+1)$-cyclic component and another vertex from
the $(q+1)$-cyclic component. A symmetry factor, $\frac{1}{2!}$,  occurs
 when $p=q$.

The right side of (\ref{FUNCTIONAL_TRIANGLE_GRAPH}) simply reflects
the constructions used to build a $(k+2)$-cyclic connected graph
In (\ref{FUNCTIONAL_TRIANGLE_MULTIGRAPH}), the term
$\frac{{\V_z}^2}{2}{W_{k,C_3}}$ represents all $(k+1)$-cyclic
multigraphs with  an ordered pair $\langle x,y \rangle$ of
marked vertices  
(see also \cite[Sect. 4, Eq. (4.2) and following]{JKLP93}). \ENDPROOF

When considering a finite set $\xi$ of forbidden configurations,
we have the following generalization of lemma \ref{LEMMA0}:
\begin{lem} \label{LEMMA_GENERAL0}
Suppose that $\xi = \{H_1, \, \cdots, H_p\}$, $|\xi| < \infty$.
Let $\gr{W}_{k+1,\xi}$, $\gr{S}_{k+1,H_i}$, $\gr{J}_{k+1,\xi}$
and $\gr{W}_{k+1,\xi}$ be the EGFs defined as in above (definition
\ref{DEF:JKXI}). Let $\rho_s$
be the finite set of all $s$-juxtapositions of member(s) of $\xi$ and
denote by $e(H_i)$ the number of edges of $H_i$.
Then, we have for graphs
\begin{eqnarray}
\V_w \gr{W}_{k+1,\xi} \, \, \, & + &  \, \, %
\sum_{H_i \in \xi}e(H_i) \gr{S}_{k+1,H_i}
\, \,  + \, \,\gr{J}_{k+1,\xi} \, \, = \, \,  %
 \, \, w \Big( \frac{{\V_z}^2 - \V_z}{2} - \V_w \Big) \gr{W}_{k,\xi} \cr
 & + & \, \, w \left(\sum_{-1\leq p \leq q \leq k+1, \, p+q=k} %
               \frac{1}{1+\delta_{p,q}} (\V_z \gr{W}_{p,\xi})%
                (\V_z \gr{W}_{q,\xi})  \right) \, .
\label{FUNCTIONAL_XI_GRAPH}
\end{eqnarray}
For the EGFs of connected $\xi$-free multigraphs, we have
\begin{eqnarray}
\V_w {W_{k+1,\xi}} \, \, \, & + &  \, \, %
\sum_{H_i \in \xi} e(H_i) {S_{k+1,H_i}}
\, \,  + \, \, J_{k+1, \xi} \, \, = \, \,  %
 \, \, w  \Big(\frac{{\V_z}^2}{2} {W_{k,\xi}}\Big) \, \, + \cr
&  & \, \, w \left(\sum_{-1\leq p \leq q \leq k+1, \, p+q=k} %
               \frac{1}{1+\delta_{p,q}} (\V_z {W_{p,\xi}})%
                (\V_z {W_{q,\xi}})  \right) \, .
\label{FUNCTIONAL_XI_MULTIGRAPH}
\end{eqnarray}
\end{lem}
(\ref{FUNCTIONAL_XI_GRAPH}) and  (\ref{FUNCTIONAL_XI_MULTIGRAPH})
are simply generalization of (\ref{FUNCTIONAL_TRIANGLE_GRAPH}) and 
 (\ref{FUNCTIONAL_TRIANGLE_MULTIGRAPH}).

\subsection{Bicyclic components without triangle}
\label{SUBSUB:BICYCLIC-TRIANGLE-FREE}
\begin{figure}[h]
\noindent
  \begin{minipage}[t]{6.5cm}
        \epsfig{file=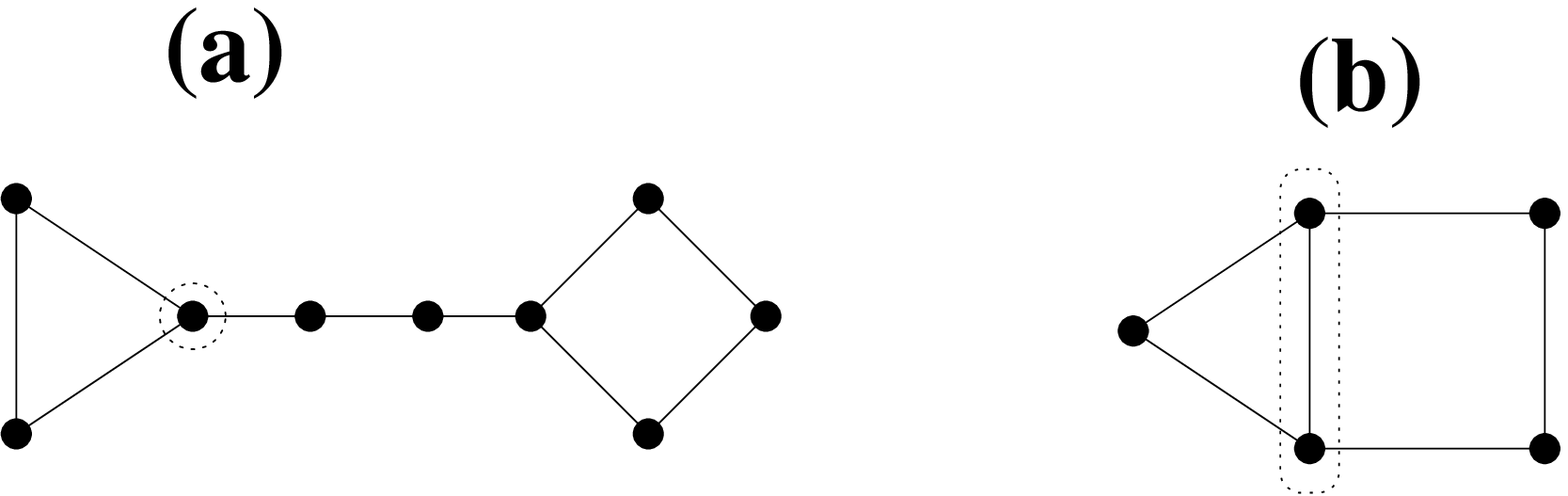,scale=0.35}
        \caption[Smooth bicyclic graphs with one occurrence of triangle.]
                {Smooth bicyclic graphs with one occurrence of triangle.}
\label{FIG:SMOOTH_S1}
     \end{minipage}
  \hfill 
  \begin{minipage}[t]{6.5cm}
     \tiny
        \epsfig{file=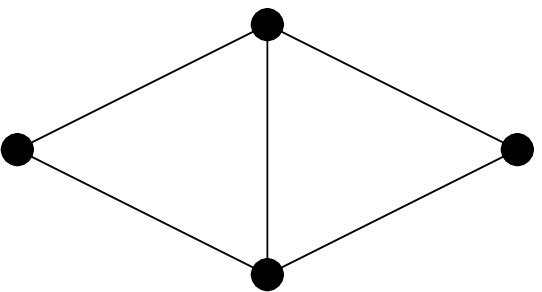,scale=0.45}
        \caption[Smooth bicyclic graph with a $1$-juxtaposition  of 2 triangles.]
                {Smooth bicyclic graph with a $1$-juxtaposition  of 2 triangles.}
\label{FIG:SMOOTH_J1} 
  \end{minipage} 
\normalsize 
\end{figure}

EGFs for respectively bicyclic graphs with one triangle 
and with exactly one juxtaposition of triangles can be
obtained using the method developed in section \ref{SEC:LINK},
 with the help of figures \ref{FIG:SMOOTH_S1} and \ref{FIG:SMOOTH_J1}. \\

\noindent  
\begin{rem} \label{rem:WRIGHT}
Since Wright's reduction method\footnote{the second method
in \cite{Wr77}, see also the proof of lemma
\ref{LEMMA_INDECOMPOSABLE} in \S \ref{SUBSUB:GENERAL-FORM}}
 suggests us to work with labelled smooth components,
figures such as \ref{FIG:SMOOTH_S1} and \ref{FIG:SMOOTH_J1}
 represent the situation after smoothing. Also for any family 
$\mathcal{F}_k$ of $(k+1)$-cyclic components with EGF
 $F_k(z)$, the EGF of smooth species of $\mathcal{F}_k$ is
simply obtained by means of substitutions of all occurrences
of $T(z)$ in $F_k(z)$ by $z$. Conversely, if $\sth{F_k}(z)$
is the EGF of smooth species of $\mathcal{F}_k$, then
$F_k(z) = \sth{F_k}(T(z))$ gives the EGF associated to the
whole family $\mathcal{F}_k$.
\end{rem} 

\begin{rem} \label{rem:TTz}
Since all EGFs we deal with can be expressed in terms of
$T(z)$ in the univariate case, and of $w$ and $T(wz)$ in the 
bivariate case, we assume that $T \equiv T(z)$ to express
univariate EGFs. In the case of bivariate
EGFs, we let $T \equiv T(wz)$. These notations should
not induce ambiguity to the reader who can read the
meaning within the context.
\end{rem}
The following figures can be used to compute the EGFs 
$\gr{S}_{1,C_3}$ and $\gr{J}_{1,C_3}$  \\

\noindent

Using similar techniques as for (\ref{DECOMPOSITION-PARTITION})
with the help of the previous figures,
we have for $\sth{\gr{S}_{1,C_3}}$ and $\sth{\gr{J}_{1,C_3}}$ 
\begin{equation}
\begin{array}{cc}
\sth{\gr{S}_{1,C_3}} (z) = & \underbrace{\frac{1}{2}z^{5}\frac{1}{1-z}}_{%
\mbox{figure \ref{FIG:SMOOTH_S1} (b)}} +  %
\underbrace{\frac{z^{6}}{4}\frac{1}{(1-z)^{2}}}_{%
\mbox{figure \ref{FIG:SMOOTH_S1} (a)}}  
\end{array}
\end{equation}
and 
\begin{equation}
\sth{\gr{J}_{1,C_3}} (z) = \frac{z^{4}}{4} \, .
\end{equation}
Again, to obtain the whole EGFs we have to substitute $z$ by $T \equiv T(z)$,
replacing all shrinked vertices of the smooth graphs by labelled rooted trees.
\begin{equation}
\gr{S}_{1,C_3}(z) = \frac{T^5}{4}\frac{(2-T)}{(1-T)^2} \, , %
\, \, \gr{J}_{1,C_3}(z) = \frac{T^4}{4} \, .
\label{EQ:S_1_C_3}
\end{equation}
Thus, using (\ref{EQ:S_1_C_3}) and (\ref{FUNCTIONAL_TRIANGLE_GRAPH})
 we have
\begin{equation}
\gr{W}_{1,C_{3}}(z)=\frac{T^{5}}{24}\frac{(2+6T-3T^{2})}{%
(1-T)^{3}}  \, .
\label{BICYCLIC-MC3FREE}
\end{equation}
We know from (\ref{TREE-POLYNOMIAL-ASYMPT}) that 
the decomposition  of formula such as
(\ref{BICYCLIC-MC3FREE}) into
sums of powers of $\frac{1}{1-T}$,
are useful in order to study the asymptotic behavior of the number of
such objects.
We have
\begin{equation}
\begin{array}{ccc}
\gr{W}_{1,C_{3}}(z) & = & \sum_{n\geq 0}\Big(\frac{5}{24}t_{n}(3)-\frac{%
25}{24}t_{n}(2)+\frac{47}{24}t_{n}(1) %
-\frac{35}{24}-\frac{5}{24}t_{n}(-1) \\ 
&  & +\frac{25}{24}t_{n}(-2)%
-\frac{5}{8}t_{n}(-3)+\frac{1}{8}t_{n}(-4)\Big) %
\frac{z^{n}}{n!} \, .
\end{array}
\label{BICYCLIC-GC3FREE-POLYNOMIAL}
\end{equation}
In order to enumerate the first multicyclic $\xi$-free components
for general $\xi$, we introduce some more techniques in the next
paragraphs.

\subsection{General techniques for first multicyclic components  %
and instantiations}
\label{SUBSUB:GRAPH-SURGERY}
In this paragraph, we give methods that can be applied to
enumerate first low-order cyclic components, i.e., with excess
$1$ and $2$ for a forbidden $p$-gon and in general for
an excess up to $l+1$ and $l+2$ for all forbidden components of excess $l$.
For e.g.,  the EGF of $C_3$-free tricyclic graphs are given
as instantiation of these methods and follows 
the formula (\ref{BICYCLIC-MC3FREE})
given above. Also, we will see
later that these techniques are useful
to obtain the forms of the EGFs $\gr{W}_{k,\xi}$ and 
$\mg{W}_{k,\xi}$ by induction (see \S \ref{SUBSUB:GENERAL-FORM}).
We consider here only connected graphs with exactly one
occurrence of $H$ since if $H^{^{\prime }}$ represents 
any juxtaposition of $H$, we can work directly in the same manner 
with a single occurrence of $H^{^{\prime }}$. 

\begin{figure}[h]
  \begin{center}
    \begin{minipage}[t]{14cm}
      \small
      \begin{center}
        \epsfig{file=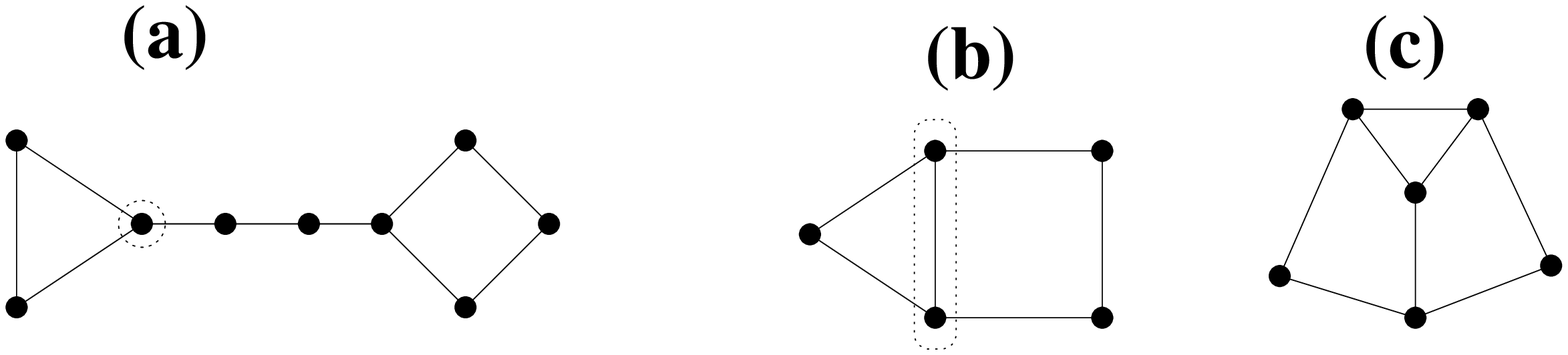,scale = 0.35}
      \end{center}
      \caption[ ] 
      { }
      \label{FIG:3}
    \end{minipage}
  \end{center}
\end{figure}

First of all, we have to prune recursively all vertices of degree $1$.
The obtained graphs are smooth. We can subdivide these graphs 
containing an occurrence of $H$ in 3 types:
types (a) and (b) are such as those represented by
figure \ref{FIG:SMOOTH_S1} and type (c) is as in the figure 
\ref{FIG:3} below where $H$ represents a triangle. 
\noindent

\noindent
The first two types (a) and (b) of figure \ref{FIG:SMOOTH_S1} 
can be described as follows: 
\begin{itemize}
\item[\rond] (a) represents the concatenation of two components $H$  
and $F$ (respectively non $H$-free and $H$-free) by a common vertex
or more generally by a path between the two components.
In the figure, $H$ is simply a triangle.  Note that a cutpoint
(a vertex whose removal increases the number of connected components)
belongs to the triangle after the recursive deletions of vertices
of degree $1$.  This is referred here as a 
\textit{serial composition} of components. 
\item[\rond] (b) is the concatenation of the same components but 
by a common edge. This construction is referred as a
\textit{parallel composition} of components.
\item[\rond] Figure \ref{FIG:3} (c) represents components which are not in 
figure \ref{FIG:SMOOTH_S1} (a) nor in figure \ref{FIG:SMOOTH_S1} (b).
\end{itemize}

\subsubsection{The serial composition or concatenation by a vertex}
\label{SUBSUB:SERIAL}
Since a graph with one cutpoint belonging to a forbidden
configuration may be considered to
be rooted at this cutpoint, the number of connected 
graphs with one cutpoint can be expressed in terms of the EGFs of the 
different subgraphs rooted at the same cutpoint 
(cf. \cite{HP73} or \cite{Selkow}). This construction
may be interpreted combinatorially as follows.

\begin{lem} \label{LEMMA7} Let $\mathcal{F}$ be a family of
 connected $H$-free graph. Denote by $\sth{F}$ the EGF 
of the graphs obtained when
smoothing a graph of $\mathcal{F}$. Let $A_1$ be the 
EGF of connected graphs containing 
 possibly many copies of $H$ and obtained
as the concatenation of graphs of $\mathcal{F}$ and of 
$H$ by a vertex belonging to $H$. Then, $A_1$ satisfies
\begin{equation}
A_1 \preceq {\left[\frac{1}{z} \, \,  \left(z\frac{\partial }{\partial z}%
\sth{F}(z)\right) \, \left( z\frac{\partial }%
{\partial z}H(z)\right)\right]}_{| z=T(z)}
  \label{PRE-ONECUTPOINT}
\end{equation}
and let $A_2$ be the EGF of all connected graphs obtained
when allowing a path starting at a vertex 
belonging to $H$ and joining 
any graph of $\mathcal{F}$. $A_2$ satisfies
\begin{equation}
A_2 \preceq {\left[\frac{1}{z} \, \left(\frac{1}{1-z}\right)%
 \,  \left(z\frac{\partial }{\partial z}%
\sth{F}(z)\right) \,  \left( z\frac{\partial }%
{\partial z}H(z)\right) \right]}_{| z=T(z)} \, .
  \label{ONECUTPOINT}
\end{equation}
In  (\ref{PRE-ONECUTPOINT}) and  (\ref{ONECUTPOINT}), equalities
hold when $H$ is two-connected.
\end{lem}

\begin{figure}
    \begin{center} \epsfig{file= 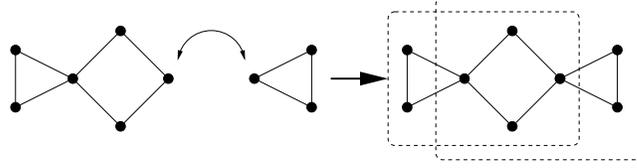,scale=0.40}
    \end{center}
    \caption[Serial composition with symmetric factor $\frac{1}{2}$]
    {Serial composition with symmetric factor $\frac{1}{2}$.}
\label{FIG:COUNTER-EXAMPLE}
\end{figure}

\noindent \textbf{Proof.} Recall that for two EGFs $A$ and $B$,
$A \preceq B$ means that $\forall n, \, 
\coeff{z^n} A(z) \leq \coeff{z^n}B(z)$ (cf. remark \ref{PRECEDENCE}). 
First, let us consider
the case where $H$ is two-connected.  In this case, the concatenation
of $H$ with a graph of $\mathcal{F}$, by a vertex of $H$, leads to a 
graph with a single copy of $H$ in the resulting graph.
Thus, the fact that there is
\textit{exactly} one occurrence of copy of $H$ in the concatenation
insures the \textit{uniqueness of the decomposition} into 
two graphs such that one belongs to $\mathcal{F}$ and the other
is (necessarily) $H$. The lemma is a combination of the
approach presented in \cite{Selkow} and Wright's reduction method 
\cite{Wr77}. We have to introduce a 
factor $\frac{1}{z}$ to relabel the common
cutpoint considered here as shared between the smooth components.
$\V_z \sth{F}(z) = z\frac{\partial}{\partial z}\sth{F}(z)$
and  $\V_z H(z) = z\frac{\partial}{\partial z}H(z)$ are used
to distinguish the vertex to be shared between
 pruned components of $\mathcal{F}$ and of $H$.  
In (\ref{ONECUTPOINT}) to
represent a possible \textit{path}, we insert the term 
$\frac{1}{1-z}$ i.e., a sequence of 
vertices of degree 2 except the two extremal nodes, between
the two sides. When substituting $z$ by $T(z)$,
 we reverse the \textit{vertexectomy} process 
starting with a smooth graph and sprout rooted trees 
from each node. Hence, in the case where $H$ is two-connected,
we have the equalities  in (\ref{PRE-ONECUTPOINT}) and  (\ref{ONECUTPOINT}).
The situation changes a bit for more general configurations.
Typically, we can have concatenations of $H$ and graphs of
$\mathcal{F}$ which can lead to a new graph with two (or more)
occurrences of $H$. This is the case depicted by figure 
\ref{FIG:COUNTER-EXAMPLE}
where $H$ is made with a triangle and a square attached by a vertex
and the graph of $\mathcal{F}$ is simply a triangle. In this special case,
we just have to introduce a symmetry factor $\frac{1}{2!}$  and then the
upper bound of (\ref{PRE-ONECUTPOINT}) is valid. In fact, the upper bound 
enumerates  graphs where  the concatenation 
such as the one obtained in figure \ref{FIG:COUNTER-EXAMPLE} are
counted twice or more.
\ENDPROOF

\subsubsection{The parallel composition or concatenation by an edge}
\label{SUBSUB:PARALLEL}
Graphs of the type represented by the figure \ref{FIG:SMOOTH_S1} (b) 
can be enumerated in a very close way.

\begin{lem} \label{LEMMA8} Let $\mathcal{F}$
and $\sth{F}$ be defined as in lemma \ref{LEMMA7} above. 
Let $B$ be the EGF associated to the graphs containing
copies of $H$ and obtained as the
concatenation of two  graphs of
$\mathcal{F}$ and of $H$ sharing a
common edge. $B$ satisfies
\begin{equation}
B \preceq {\left[ \frac{2}{wz^{2}} \, \, %
\left(w \frac{\partial }{\partial w}\sth{F}(w,z)\right) \, \, %
\left(w \frac{\partial }{\partial w} %
H(w,z)\right)\right]}_{| wz=T(wz)} \, .
\label{ONECUTEDGE}
\end{equation}
\end{lem}

\noindent \textbf{Proof.} The formula  (\ref{ONECUTEDGE})
differs slightly from the one in (\ref{ONECUTPOINT}). The 
factor $\frac{2}{wz^{2}}$ comes from
the fact that we have here, as in the figure \ref{FIG:SMOOTH_S1} (b),
 a common edge which is defined by his two common vertices 
and can be seen as a root-edge. A graph such as those represented by the 
figure \ref{FIG:SMOOTH_S1} (b) can be considered as pendant
to this edge. Also, we have the equality whenever
$H$ is two-connected. Otherwise symmetries can arise but the
upper bound of  (\ref{ONECUTEDGE}) remains valid for the same reasons
as for  (\ref{PRE-ONECUTPOINT}) and  (\ref{ONECUTPOINT}). \ENDPROOF

Unfortunately, equation likes (\ref{FUNCTIONAL_XI_GRAPH})
of lemma  \ref{LEMMA_GENERAL0} are much
easier to propose than to really solve. However, we can derive 
the EGF of the first multicyclic $H$-free components 
by applying the techniques presented above. 


\subsubsection{The example of triangle-free graphs}
\begin{figure}[h]
  \begin{center}
    \begin{minipage}[t]{12cm}
      \small
      \begin{center}
        \epsfig{file= 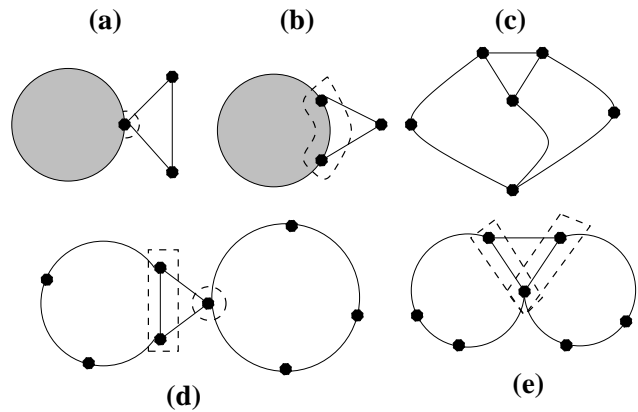,scale= 0.5}
      \end{center}
    \caption{Basic tricyclic graphs with exactly one triangle. The
      subgraph in grey represent bicyclic triangle-free components.}
    \label{FIG:solo2C3}
  \end{minipage}
  \end{center}
\end{figure}

\begin{figure}[h]
  \begin{center}
    \begin{minipage}[t]{12cm}
      \small
      \begin{center}
       \epsfig{file= 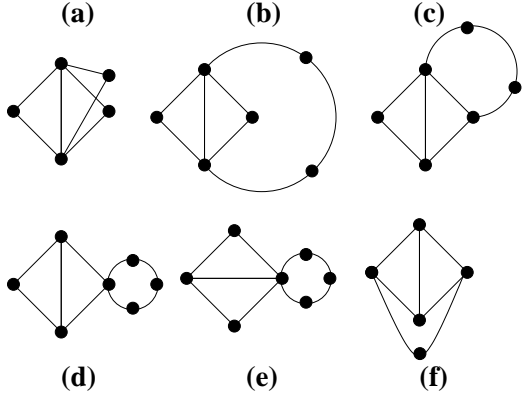,scale=0.5}
     \end{center}
       \caption{Basic tricyclic graphs with juxtapositions of triangles.}
     \label{FIG:JUXTA2C3}
     \end{minipage}  
   \end{center}
\end{figure}

The EGFs of unicyclic and bicyclic graphs without triangles 
are given by formulae (\ref{ACYCLIC-GC3FREE}) 
and (\ref{BICYCLIC-MC3FREE}).
For graphs having $2$ excesses, the removal of all edges and
vertices by the Wright's reduction method leads to the set of graphs
represented by figure \ref{FIG:solo2C3} for graphs 
containing $1$ triangle and
figure \ref{FIG:JUXTA2C3}  for graphs with a juxtaposition of triangles.
\noindent
 
\noindent 
As before, given a family $\mathcal{F}$ of graphs, we denote by
$\sth{F}$ the EGF of \textit{smooth} elements of $\mathcal{F}$,
i.e., graphs without endvertices (vertices of degree $1$). 
The bivariate EGF of bicyclic triangle-free smooth graphs,
$\sth{\gr{W}_{1,C_3}}$ is obtained from (\ref{BICYCLIC-MC3FREE}), namely
\begin{equation}
\sth{\gr{W}_{1,C_3}}(w,z)=w\frac{w^{5}z^{5}}{24} %
\frac{(2+6wz-3w^{2}z^{2})}{(1-wz)^{3}}
\label{eqn:FreeC3-1-smooth}
\end{equation}
Note that $\V_w C_3(w,z)= \V_z C_3(w,z) = \frac{w^3z^3}{2}$.
Thus, the application of the lemmas \ref{LEMMA7} and  
\ref{LEMMA8} to the smooth graphs depicted by
figures \ref{FIG:solo2C3} (a) and
\ref{FIG:solo2C3} (b) gives 
\begin{equation}
{\frac{w^{3}z^{2}}{2(1-wz)} \, \, \V_z  %
\big(\sth{\gr{W}_{1,C_3}}(w,z)\big) %
+ w^{2}z \, \, \V_w\big(\sth{\gr{W}_{1,C_{3}}}(w,z)\big) } \, .
\label{eqn:SoloC3-2-lemma78}
\end{equation}
Similarly, we have for smooth graphs represented by the figure 
\ref{FIG:solo2C3} (d)
\begin{equation} 
{\frac{1}{z(1-wz)}  \Big(\frac{2}{wz^2} \, %
\big(\V_w \sth{\gr{W}_{0,C_3}}(w,z) \big) \, \big(\frac{w^3z^3}{2}%
 \Big) \Big) %
\Big( \V_z \sth{\gr{W}_{0,C_3}}(w,z) \Big)} 
\label{eqn:solo2C3-d}
\end{equation}
and for figure \ref{FIG:solo2C3} (e), we find
\begin{equation}
\frac{2}{wz^2}  %
\left( \frac{2}{wz^2} \Big( \frac{w^3z^3}{2} \Big) %
\Big( \V_w \sth{\gr{W}_{0,C_3}}(w,z) \Big)^2 \right) \, .
\label{eqn:solo2C3-e}
\end{equation}
A simple way to enumerate the smooth graphs represented by
the figure \ref{FIG:solo2C3} (c) is to consider
that the three paths between the triangle and the vertex
$v$ are symmetric. Taking into account the fact
that only one of these three paths can be reduced
to a simple edge (to avoid another triangle), we have
the following EGF associated to these smooth graphs
\begin{equation}
\frac{z^7}{3!(1-z)^3} + \frac{z^6}{2!(1-z)^2} \, .
\label{eqn:solo2C3-c}
\end{equation}
In total, the bivariate EGF for all graphs such that smooth species are
depicted by the figures \ref{FIG:solo2C3} (c), \ref{FIG:solo2C3} (d) and
\ref{FIG:solo2C3} (e) is given by
\begin{equation}
w^2 \frac{T^6}{6}\frac{(3-2T)}{(1-T)^3} + \frac{w^2T^7}{%
2(1-T)^2} + \frac{w^2T^8}{4(1-T)^3} \, .
\label{eqn:SoloC3-2-others}
\end{equation}
Summing (\ref{eqn:SoloC3-2-lemma78}) and (\ref{eqn:SoloC3-2-others}%
), one can deduce the bivariate EGF for tricyclic graphs 
containing exactly a triangle
\begin{equation}
\gr{S}_{2,C_3}(w,z) = \frac{w^2T^6}{48} %
       \frac{(48+18T-140T^2+119T^3-30T^4)}{(1-T)^5} \, .
\label{eqn:SoloC3-2}
\end{equation}

We turn now to the enumeration of tricyclic graphs
with one occurrence of juxtaposition of triangles.
The figure \ref{FIG:JUXTA2C3}  represents the 
2-excess smooth graphs with juxtapositions of
triangles. 

\noindent
We observe that figures \ref{FIG:JUXTA2C3} (b) and  \ref{FIG:JUXTA2C3} (c)
can be handled with the techniques of lemma \ref{LEMMA8} using the EGF
$\sth{\gr{W}_{0,C_3}}$ and $\frac{w^5z^4}{2!2!}$ (which is the EGF of the
smooth juxtaposition of $2$ triangles).
Similarly, we can use lemma \ref{LEMMA7} for the figures
  \ref{FIG:JUXTA2C3} (d) and  \ref{FIG:JUXTA2C3} (e).
The EGF associated to the smooth graph of figure  \ref{FIG:JUXTA2C3} (a)
is simply $\frac{w^7z^5}{2!3!}$, and the one for
smooth graphs depicted by the figure  \ref{FIG:JUXTA2C3} (f)
is $\frac{w^7z^5}{4(1-wz)}$. In fact, graphs such as 
the one drawn in figure  \ref{FIG:JUXTA2C3} (f) can be obtained
by replacing an edge of the complete graph $K_4$ with
a path of length at least $2$. 
The EGF that corresponds to the figure \ref{FIG:JUXTA2C3}  is 
then
\begin{eqnarray}
 & & \sth{\gr{J}_{2,C_3}}(w,z)  = \frac{w}{z(1-wz)} %
\V_z\big(\frac{w^5z^4}{4}\big) \V_z\big(\sth{\gr{W}_{0,C_3}}(w,z)\big) \cr
 & & +  \frac{2}{wz^2} \V_w\big(\frac{w^5z^4}{4}\big) %
 \V_w\big(\sth{\gr{W}_{0,C_3}}(w,z)\big) %
 +  {w^2 \frac{(wz)^5}{2!3!}} +  {w^2\frac{(wz)^5}{4(1-wz)}} \, .
\end{eqnarray} 
Thus, the bivariate EGF of tricyclic graphs containing
exactly a juxtaposition of triangles is
\begin{equation}
\gr{J}_{2,C_3}(w,z) = \frac{w^2T^5}{6}%
\frac{(2+5T-4T^2)}{(1-T)^2} \, .
\label{eqn:JuxtaC3-2}
\end{equation}
The bivariate EGF of tricyclic triangle-free graphs is then obtained using
(\ref{eqn:SoloC3-2}), (\ref{eqn:JuxtaC3-2}) 
and (\ref{FUNCTIONAL_TRIANGLE_GRAPH}), namely,
\begin{equation} 
 \gr{W}_{2,C_3}(w,z)=w^2\frac{T^6}{48} %
 \frac{(7+36T-18T^2-40T^3+40T^4-10T^5)}{(1-T)^6} \, .
\label{eqn:FreeC3-2}
\end{equation}

\subsection{General forms of the EGFs of $\xi$-free components}
\label{SUBSUB:GENERAL-FORM}
Although lemmas \ref{LEMMA0} and \ref{LEMMA_GENERAL0} do not allow us 
to solve completely the problems of 
enumerating $\xi$-free connected graphs with
a given number of vertices and edges, the combination of
these lemmas with subtle combinatorial
constructions provides alternative solutions to get the general forms
of the EGFs $\gr{W}_{k,\xi}$ and ${W_{k,\xi}}$. Recall the following
theorem due to \aut{Wright}
\begin{thm}[Wright 1977] \label{TH_WRIGHT_FORM} 
For $k \geq 1$, the EGFs, $\gr{W_{k}}$,
 of $(k+1)$-cyclic graphs can be expressed as a finite sum
of powers of $\frac{1}{1-T(z)}$ with rational coefficients and
we have
\begin{equation}
\gr{W}_{k}(z) = \frac{b_k}{(1-T(z))^{3k}} - \frac{c_k}{(1-T(z))^{3k-1}} %
+ \sum_{ 2 \leq s \leq 3k-2} \frac{\omega_{k,s}}{(1-T(z))^{s}} \, .
\label{EQ:TH_WRIGHT_FORM}
\end{equation}
The $(b_k)_{k \geq 1}$ are called
the Wright's constants of first order (also called 
Wright-Louchard-Tak\'acs constants, see for e.g. \cite{Sp97}).
 $b_1=\frac{5}{24}$ and for $k \geq 1$, $b_k$ is defined recursively by
\begin{equation}
2 (k+1) b_{k+1} = 3k(k+1)b_k+ 3 \sum_{t=1}^{k-1} t(k-t)b_t b_{k-t} \, .
\label{EQ:B_K}
\end{equation}
The $(c_k)_{k \geq 1}$ are the Wright's constants of
second order and are defined recursively, using (\ref{EQ:B_K}), by
$c_1 = \frac{19}{24}$ and for $k \geq 1$
\begin{eqnarray}
 2 (3k+2) c_{k+1} &=& 8 (k+1) b_{k+1} + 3k b_k + (3k+2)(3k-1) c_k \cr
 & + & 6\sum_{t=1}^{k-1} t (3k - 3t -1) b_t c_{k-t} \, .
\label{EQ:C_K}
\end{eqnarray}
\end{thm} 
The proof of theorem \ref{TH_WRIGHT_FORM} is an interesting 
combinatorial exercise involving essentially the pointing
operators $\V_w$ and $\V_z$ (see \cite{Wr77, JKLP93}). 
Note that formulae (\ref{EQ:TH_WRIGHT_FORM}), (\ref{EQ:B_K}) and
(\ref{EQ:C_K}) are obtained with \aut{Wright}'s fundamental
differential recurrence (well explained in \cite[section 6]{JKLP93})
and which is written here with the notations of this paper
\begin{eqnarray}
\V_w \gr{W}_{k+1} &=& \, \,  %
 \, \, w \Big( \frac{{\V_z}^2 - \V_z}{2} - \V_w \Big) \gr{W}_{k} \cr
 & + & \, \, w \left(%
\sum_{ -1\leq p \leq q \leq k+1, \, p+q=k} %
               \frac{1}{1+\delta_{p,q}} (\V_z \gr{W}_{p})%
                (\V_z \gr{W}_{q}) \right) \, .
\label{FUNCTIONAL_WRIGHT}
\end{eqnarray}

For our connected $(k+1)$-cyclic triangle-free graphs, we have the
following existence theorem on the forms of their EGFs:
\begin{thm} \label{THM_GENERAL_FORM}
There exists rational $\omega_{k,i}^{(C_3)}$ such that for all
 $k \geq 2$, the univariate EGF, $\gr{W}_{k,C_3}$, associated 
to $(k+1)$-cyclic triangle-free graphs,
 is of the form:
\begin{equation}
\gr{W}_{k,C_3}(z) = \frac{b_k}{( 1-T)^{3k}} - %
\frac{ \cpt{k} } %
{ (1-T )^{3k-1}} + %
\sum_{i \leq 3k-2} %
 \frac{ {\omega}_{k,i}^{(C_3)} }{(1-T)^{i}} 
\label{PWK}
\end{equation} 
where  $T\equiv T(z)$, the summation is finite and
the coefficients $\cpt{k}$ are defined, for all $k \geq 1$, by
\begin{equation}
\begin{array}\{{rl}.
           &  \cpt{1}=\frac{25}{24} \, , \\
           &  \cpt{k+1} = c_{k+1} + \frac{3}{2}k b_k \, .
\end{array}
\label{CPRIME_0}
\end{equation}
\end{thm}
Before proving theorem \ref{THM_GENERAL_FORM},
the connected components with one occurrence of triangle
are subdivided into $3$ kinds of constructions, according
to the degrees of the vertices of the unique triangle (after smoothing). 
Let us define these classifications. A smooth graph containing
a triangle is of three kinds:
\begin{itemize}
\item[-] exactly one vertex of the triangle is of degree $ \geq 3$,
\item[-] exactly two vertices of the triangle are of degree $\geq 3$,
\item[-] the $3$ vertices of the triangle are all of degree $\geq 3$.
\end{itemize}

\noindent
\begin{figure}[h]
\noindent
  \begin{minipage}[t]{4.0cm}
        \epsfig{file=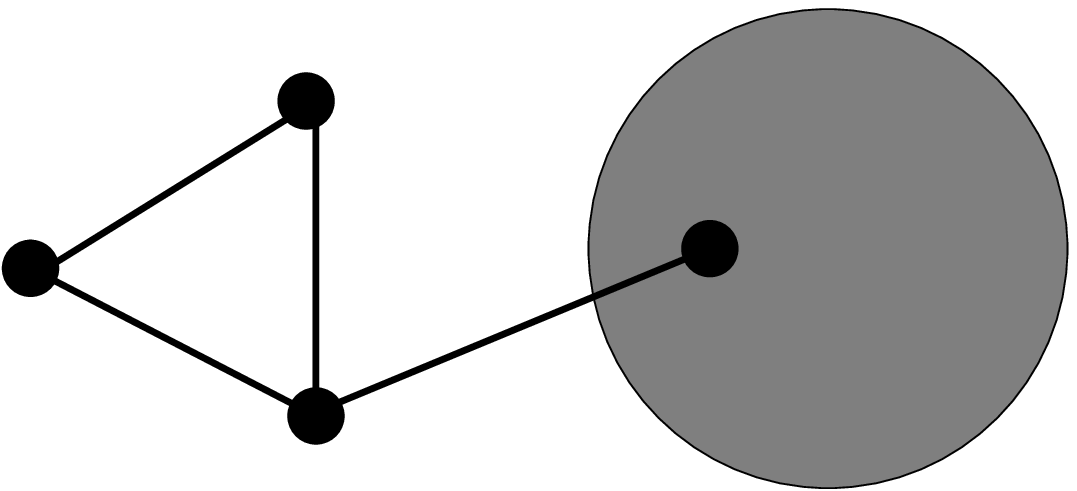,scale=0.25}
        \caption[Type I]
                {One vertex of the triangle
                 is of degree $\geq 3$.}
\label{FIG:SOLO1}
     \end{minipage}
  \hfill 
  \begin{minipage}[t]{4.0cm}
     \tiny
        \epsfig{file=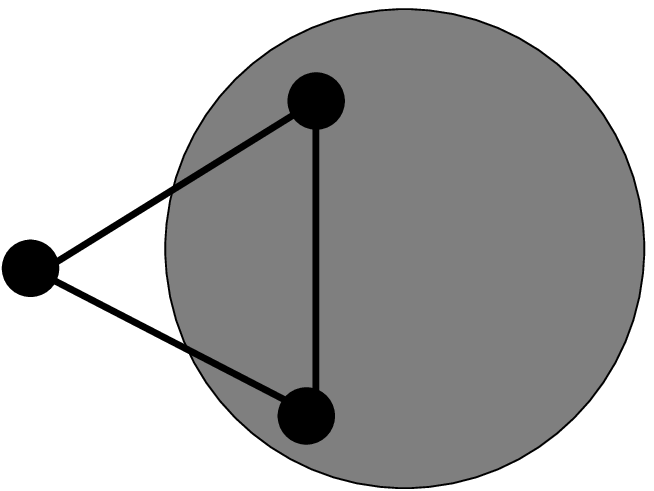,scale=0.25}
        \caption[Type II]
                {Two vertices of the triangle
                 are of degree $\geq 3$.}
                 
\label{FIG:SOLO2}
  \end{minipage}
  \hfill 
  \begin{minipage}[t]{4.0cm}
     \tiny
        \epsfig{file=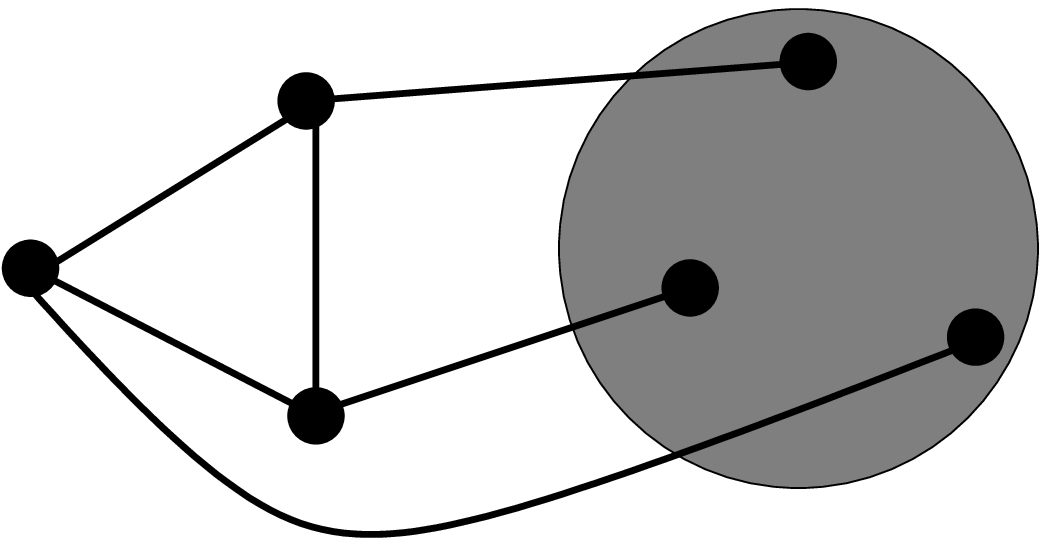,scale=0.25}
        \caption[Type III]
        {Other smooth components.}
        \label{FIG:SOLO3}
  \end{minipage}
\normalsize 
\end{figure}

\noindent
Graphs whose situations after smoothing
are depicted by figures \ref{FIG:SOLO1} and \ref{FIG:SOLO2}
can be handled by the techniques of lemmas \ref{LEMMA7}
and \ref{LEMMA8}, and will be considered more
precisely later. Note that in the figures, the right 
parts (in grey) of the constructions 
correspond to multicyclic structures without triangle.
The lemma \ref{LEMMA_INDECOMPOSABLE}
gives the form of the EGF of the connected component with 
exactly one occurrence of
triangle depicted by the figure \ref{FIG:SOLO3}.
\begin{lem} \label{LEMMA_INDECOMPOSABLE}
The EGF of $(k+1)$-cyclic graphs containing one occurrence of
triangle with all of its vertices  of degree at least $3$ has the
following form
\begin{equation}
\sum_{s \leq 3k-3} \frac{\epsilon_{k,s}}{(1-T(z))^{s}} \, 
\end{equation}
where the summation is finite and the coefficients $\epsilon_{k,s}$
are rational numbers.
\end{lem}

\noindent \textbf{Proof.} Our idea is to apply 
Wright's reduction method on our specific configuration. 
Since this method is known but is not that
familiar, we repeat here the main steps. Suppose that 
we have a connected graph with $k$ edges more than vertices containing one triangle
and suppose that the recursive suppressions of vertices of degree $1$ lead
to a graph of the type depicted by figure \ref{FIG:SOLO3}.
That is, the obtained smooth graph has $t$ vertices of
degree at least $2$ and $t+k$ edges (here, $t$ is less that or
equal to the number of vertices of the original graph).
This way, we get a smooth graph with $r$ vertices of degree
at least $3$, $r \leq 2k$. These vertices of degree $\geq 3$
are called \textit{special vertices} and let us \textit{color}
the edges of the triangle in order to distinguish them.
 The paths between these points, except the colored edges of the
triangle, are of four kinds and we apply the following special
operations on them (see \cite[Sect.~ 6]{Wr77}):
\begin{itemize}
\item[1.] An $\alpha$-path begins and ends with the same special point
and so must have at least two interior points. We elide all its interior
points except two of them.
\item[2.] A $\beta$-path joins two different special vertices 
and we elide all its interior points.
\item[3.] If two different special vertices are joined by more than
one special path, at most one of these paths is reduced to a single edge
which we call a $\delta-path$.
\item[4.] The remaining paths, or all the paths if there is
no $\delta$-path, are called $\gamma$-paths and for each
$\gamma$-path, we elide all its interior points except one of them.
\end{itemize}
The obtained graph is called \textit{Wright's basic graph}.
Denote respectively by $a$, $b$, $c$ and $d$ 
the number of $\alpha$-, $\beta$-, $\gamma$-
 and $\delta$- paths. Since each elision has removed exactly 
one edge and one vertex, the number of vertices of the 
basic graph is exactly $r+2a+c$. Taking into account,
the colored edges of the triangle and the operations made upon
the special paths, the number of edges in the basic graph is
$r+2a+c+k = 3a+b+2c+d+3$. Thus, we have $a+b+c+d+3 = r+k \leq 3k$.
We find
\begin{equation}
 a+b+c \leq 3k -3 \, .
\label{ELIDEELIDE}
\end{equation}
To obtain any of the original graphs without vertices of degree $1$,
we distribute the previously $t-r-2a-c$ elided  nodes
on the $\alpha$-, $\beta$- and $\gamma$- paths.  (\ref{ELIDEELIDE}) gives
us ideas on the number of ways to redistribute these points:
suppose that $f(n)$ is the number of labelings of the $(n,n+k)$-graphs
which can produce the considered basic graph. Let $F(z)$ be their EGF:
\begin{equation}
F(z) = \sum_n f(n) \frac{z^n}{n!} \, .
\end{equation} 
To obtain each of the original $(t,t+k)$ graphs without endvertices,
the distribution of the $(t-r-2a-c)$ nodes on the
$(a+b+c)$ $\alpha$-, $\beta$- and $\gamma$-paths can be done
in $y$ ways where $y$ is the number of
partitions of $(t-r-2a-c)$ into $(a+b+c)$ parts.
Relabel the obtained graph and replace the $t$ vertices with
$t$ rooted and labelled trees. 
All the graphs are 
enumerated but they are not all different. In fact, they
are enumerated $g$ times where $g$ is the order of
the automorphisms of the current Wright's basic graph. 
Thus, we have
\begin{equation}
g F(z) = \sum_t y T(z)^t = \frac{T(z)^{r+2a+c}}{(1-T(z))^{a+b+c}} \, .
\end{equation}
Summing over all the finitely many possible basic graphs,
we obtain the lemma.
\ENDPROOF

\noindent \textbf{Proof of theorem \ref{THM_GENERAL_FORM}.}  Denote by
\PW{k}, \PS{k} and \PJ{k} the following properties:
\begin{itemize}
\item[\rond] \PW{k}~: $\gr{W}_{k,C_3}$ is of the form
given by the equation (\ref{PWK}).
\item[\rond] \PS{k}~: \\
If $k=1$,
\begin{equation}
\gr{S}_{1,C_3}(z)  =  %
\frac{1}{4\, (1-T)^{2}}-\frac{1}{(1-T)}-\frac{1}{4}{T}^{4}+%
\frac{1}{4}{T}^{2}+\frac{1}{2}T +\frac{3}{4} \, 
\end{equation}
and for all $k \geq 2$, $\gr{S}_{k,C3}$ is of the form
\begin{equation}
\gr{S}_{k,C_3}(z) = \frac{3 (k-1) b_{k-1}}{2 \, \big( 1-T(z)  \big)^{3k-1}} %
+ \sum_{i \leq 3k-2} %
\frac{\sigma_{k,i}^{(C_3)}}{\big( 1-T(z)  \big)^{i}} \, .
\label{PSK}
\end{equation}
\item[\rond] \PJ{k}~: \\
If $k=1$
\begin{equation}
 \gr{J}_{1,C_3}(z)  =  \frac{T^4}{4} \, 
\end{equation}
and if $k=2$, we have
\begin{equation}
\gr{J}_{2,C_3}(z)  =  \frac{1}{2 (1-T)^2 }- %
\frac{2}{(1-T)} + \frac{3}{2} + T + \frac{T^2}{2} %
- \frac{T^4}{2} - \frac{2 T^5}{3}\, .
\label{DECOMPOSITION_J2}
\end{equation}
For all $k \geq 3$, $\gr{J}_{k,C_3}$ is of the form
\begin{equation}
\gr{J}_{k,C_3}(z) = \frac{3 (k-2) b_{k-2}}{\big( 1-T(z) \big)^{3k-4}} %
+ \sum_{i \leq 3k-5} %
\frac{\upsilon_{k,i}^{(C_3)}}{\big( 1-T(z) \big)^{i}} \, .
\label{PJK}
\end{equation}
\end{itemize}
where the coefficients $(\omega_{k,i}^{(C_3)})$, $(\sigma_{k,i}^{(C_3)})$
and $(\upsilon_{k,i}^{(C_3)})$ are rational numbers and the summations
in (\ref{PWK}), (\ref{PSK}) and (\ref{PJK}) are \textbf{finite}.

We will show by \textit{induction} on  $k$, that for all $k \geq 1$,
the properties \PW{k}, \PS{k} and \PJ{k} described above
are simultaneously verified. To do this,
we have
\PW{1}, \PS{1}, \PJ{1} and \PJ{2}  
and we have to check that if
\PW{i}, \PS{i} and \PJ{i} are true for all $i$ such that
$ 1 \leq i \leq k-1$ then
\PW{k}, \PS{k} and \PJ{k} are also satisfied.
Note that due to the presence of the factor $(k-1)$ in 
(\ref{PSK}), resp. $(k-2)$ in (\ref{PJK}), 
we have to give $\gr{S}_{1,C_3}$, $\gr{J}_{1,C_3}$
and $\gr{J}_{2,C_3}$.
Rewriting (\ref{eqn:FreeC3-2}) and  (\ref{eqn:SoloC3-2})
as sums of powers of $\frac{1}{1-T}$, we have
\begin{eqnarray}
& & \gr{W}_{2,C_3}(z) =  \frac{5}{16 (1-T)^6}   %
 - \frac{5}{3 (1-T)^5} + \frac{167}{48 (1-T)^4 }%
- \frac{91}{24 (1-T)^3}  \cr
& & + \frac{55}{16 (1-T)^2 } %
- \frac{35}{8 (1-T) } + \frac{125}{48} %
+ \frac{17 T}{12} + \frac{11 T^2}{24} - \frac{5 T^3}{24}  %
- \frac{5 T^4}{12} - \frac{5 T^5}{24} \, , \cr
& & \gr{S}_{2,C_3}(z)  =  \frac{5}{16 (1-T)^5 } %
- \frac{65}{48 (1-T)^4} + \frac{7}{3 (1-T)^3} %
  - \frac{73}{24 (1-T)^2}  \cr
 & & + \frac{61}{12 (1-T)} %
- \frac{10}{3} - \frac{103 T}{48} - \frac{53 T^2}{48} %
- \frac{5 T^3}{48} %
 + \frac{31 T^4}{48} + \frac{5 T^5}{8}  \, .
\end{eqnarray}
Thus, $\gr{S}_{2,C_3}(z)$, $\gr{J}_{2,C_3}(z)$,
and $\gr{W}_{2,C_{3}}(z)$ can be formulated as finite sums
of power of $\frac{1}{(1-T)}$ and properties \PW{2}, \PS{2} and \PJ{2} 
are satisfied. Note that  we let $b_0 = \frac{1}{2}$, due to the fact that
$\V_z \, \gr{W}_{0,C_3}(z) = \frac{1}{2} \, \frac{T^4}{(1-T)}$.
Now, suppose that \PW{i}, \PS{i} and \PJ{i} are true for 
$i \in \coeff{1,\, k-1}$. If we want to compute directly
$\gr{W}_{k,C_3}$, the differential recurrence relation
(\ref{FUNCTIONAL_TRIANGLE_GRAPH}) of
lemma \ref{LEMMA0} is not useful except if we know 
the EGFs $\gr{S}_{k,C_3}$ and $\gr{J}_{k,C_3}$. However, assuming
that \PW{i}, \PS{i} and \PJ{i} are true for 
$i \in \coeff{2,\, k-1}$, we can compute the forms of $\gr{S}_{k,C_3}$ and
$\gr{J}_{k,C_3}$ using combinatorial decompositions of these
graphs. In the rest of this proof, our attention will be
focused on the terms involving $\frac{1}{(1-T(z))^{3k}}$
and $\frac{1}{(1-T(z))^{3k-1}}$ for
$\gr{W}_{k,C_3}$ and $\frac{1}{(1-T(z))^{3k-1}}$ 
 for $\gr{S}_{k,C_3}$.
Under the hypothesis of the induction, let us compute
the forms of $\gr{S}_{k,C_3}$ and $\gr{J}_{k,C_3}$.
More specifically, the components represented by figures \ref{FIG:SOLO1} and
\ref{FIG:SOLO2} can be decomposed and the forms of their EGFs can be
computed using the EGF of the triangle
(eq. (\ref{TRIANGLE-TRIANGLE-EGF})), the operator $\V_z$ (to distinguish
the  common point) and the form of the EGF $\gr{W}_{k-1,C_3}$ which is 
assumed by the induction hypothesis. 
Recall that $\sth{\gr{W}_{k-1,C_3}}$ denotes the EGF of
\textit{$k$-cyclic smooth graphs} without triangle obtained
when deleting recursively all vertices of degree $1$.
Using lemma \ref{LEMMA7}, we obtain the univariate 
EGF of all the graphs such that the 
situation after smoothing is depicted by figure \ref{FIG:SOLO1}, namely
\begin{equation}
{\left[ \frac{1}{z} \, \frac{1}{1-z} \, %
\V_z (\frac{z^3}{3!}) \V_z \sth{\gr{W}_{k-1,C_3}}(z) \right]}_{| z=T(z)}
\end{equation}
Similarly, the smooth graph represented by figure \ref{FIG:SOLO2} can be
enumerated using the operator $\V_w$. We obtain the following
bivariate EGF 
\begin{equation}
 {\left[ \frac{2}{wz^2}  \V_w \big( \frac{w^3 z^3}{3!} \big) %
\V_w \big( \sth{\gr{W}_{k-1,C_3}}(w,z)\big)  \right]}_{| wz = T(wz)} \, 
\end{equation}
Using the form of the EGF of $(k+1)$-cyclic components given
by lemma \ref{LEMMA_INDECOMPOSABLE}, we find the form of the bivariate 
EGF of smooth graphs of $\sth{\gr{S}_{k,C_3}}$,
\begin{eqnarray}
 & & \sth{\gr{S}_{k,C_3}}(w,z) = \left( \frac {w} {z(1-wz)} %
 \V_z \big( \frac {w^3 z^3} {3!} \big) %
\V_z \big( \sth{\gr{W}_{k-1,C_3}} (w,z) \big)  \right) \cr
 & & + \left( \frac{2}{wz^2}  \V_w \big( \frac{w^3 z^3}{3!} \big) %
\V_w \big( \sth{\gr{W}_{k-1,C_3}}(w,z)\big)  \right) %
 + w^k \sum_{i \leq 3k-2} %
\frac{ s_{k,i}^{(C_3)} }{\big( 1-wz \big)^i} \, . 
\label{SMOOTH_S}
\end{eqnarray} 
Remark that the constants $s_{k,i}^{(C_3)}$ are not those described by
eq. (\ref{PSK}) because we have to take into account the terms
from $\frac{2}{wz^2}  \V_w \big( \frac{w^3 z^3}{3!} \big)
\V_w \big( \sth{\gr{W}_{k-1,C_3}}(w,z)\big)$.
Thus, we find
\begin{eqnarray}
 & & \sth{\gr{S}_{k,C_3}}(w,z)  \, \, =  \frac{w^3z^2}{2(1-wz)} \times %
  w^{k-1} \, \V_z \left( \frac{b_{k-1}}{(1-wz)^{3k-3}} 
+ \sum_{i \leq 3k-4}%
 \frac{s_{k-1,i}^{(C_3)}}{(1-wz)^i} \right)  \cr
& &  + w^2 z \V_w \left(\frac{w^{k-1}b_{k-1}}{(1-wz)^{3k-3}} %
 + \sum_{i \leq 3k-4}%
 \frac{w^{k-1} s_{k-1,i}^{(C_3)}}{(1-wz)^i} \right)  %
   + w^k \sum_{i \leq 3k-2} %
\frac{ s_{k,i}^{(C_3)} }{\big( 1-wz \big)^i} \, . 
\end{eqnarray}
A bit of calculus leads to the EGF of $(k+1)$-cyclic
components with exactly one triangle
\begin{equation}
\gr{S}_{k,C_3}(w,z) = w^k \left( %
\frac{3 (k-1) b_{k-1}}{2 \big( 1-T  \big)^{3k-1}} %
+ \sum_{i \leq 3k-2} %
\frac{\sigma_{k,i}^{(C_3)}}{\big( 1-T  \big)^{i}} \right) \, .
\label{FORME_S_C3}
\end{equation}
and \PS{k} is verified. Similarly,
the same principles can be used to compute
the form of $\gr{J}_{k,C_3}$ when replacing the  single occurrence of
triangle by a single occurrence of juxtaposition of triangles which can 
be considered in its turn as a single subgraph. For this purpose,
we have to replace the EGF $\frac{w^3z^3}{3!}$ of the triangle
by EGFs of juxtapositions of triangles, viz.
 $\frac{w^5z^4}{2!2!}$ (EGF of the smooth graph 
depicted by figure \ref{FIG:SMOOTH_J1}),
  $\frac{w^7z^5}{2!3!}$, $\cdots$, 
$\frac{w^{2i+1}z^{i+2}}{2!i!}$, $\cdots$ . We find
\begin{eqnarray}
 & & \sth{\gr{J}_{k,C_3}} (w,z) \, \,=   \frac {w} {z(1-wz)}%
 \V_z\big( \frac {w^5 z^4} {4} \big) \, %
\V_z \big( \sth{\gr{W}_{k-2,C_3}} (w,z) \big)  \cr
& &  \, + \, \frac {w} {z(1-wz)} %
 \V_z\big( \frac {w^7 z^5} {12} \big) %
\V_z \big( \sth{\gr{W}_{k-3,C_3}} (w,z) \big) \cr
 & & + \frac{2}{wz^2}  \V_w \big( \frac{w^5 z^4}{4} \big) %
\V_w \big( \sth{\gr{W}_{k-2,C_3}}(w,z)\big) %
 + w^k \sum_{i \leq 3k-3} %
\frac{ \iota_{k,i}^{(C_3)} }{\big( 1-wz) \big)^i} \, . 
\end{eqnarray}
Hence, we have the form of $3\gr{S}_{k,C_3}+\gr{J}_{k,C_3}$
which starts with $\frac{9(k-1)b_{k-1}}{2 (1-T)^{3k-1}}$.
We need some useful notations, mainly related to those of Wright
 \cite{Wr77,Wr80}.
Denote by $\xx$ the following EGF
\begin{equation}
\xx \equiv 1-T \, .
\label{eq:DEFTHETA}
\end{equation}
Let $\Lambdat{1} = 0$ and for all $k \geq 2$, let
$\Lambdat{k}$ be the following formal power series 
\begin{equation}
\Lambdat{k}: \,  \Lambdat{k}(z) = \sum_{t=1}^{k-1} %
\Big(\V_z \gr{W}_{t,C_3}(z) \Big) %
\Big(\V_z \gr{W}_{k-t,C_3}(z)\Big) \, .
\label{EQ:LAMBDA_K}
\end{equation}
Let $F$ be an EGF. For all $k \geq 1$, 
we denote by $\Delta$ and $\Omegat{k}$ the following operators
\begin{equation}
\Delta_{k+1}: \, \, \Delta_{k+1} \, \big(F \big) %
= 2 \Big( k+1 - T \frac{\partial}{\partial T}\Big) \, \big(F\big) \, 
\label{OP:DELTA_K}
\end{equation}
and
\begin{equation}
\Omegat{k}: \, \, \Omegat{k} \, \big(F\big) =%
 \Big( \big( \V_z^2 - 3 \V_z - 2k\big) +%
 2 \big( \V_z \gr{W}_{0,C_3}(z) \big) \V_z \Big) \, \big(F \big) \, .
\label{OP:OMEGAT_K}
\end{equation}
Using these notations, we remark that the functional equation
(\ref{FUNCTIONAL_TRIANGLE_GRAPH}) of lemma \ref{LEMMA0}
can be reformulated as follows
\begin{eqnarray}
& & \Delta_{k+1} \gr{W}_{k+1,C_3} %
+ 6 \gr{S}_{k+1,C_3} + 2 \gr{J}_{k+1,C_3} = \cr
& & \, \, \, \, %
\Omegat{k} \gr{W}_{k,C_3} + \Lambdat{k} \, , \, \, (k \geq 1) \, .
\label{AVEC_OP_C3} 
\end{eqnarray}
Then, we remark that
\begin{equation}
\Delta_{k} {\xx}^{-t} = \Delta_{k} \frac{1}{(1-T)^{t}} = %
2 \xx^{-t} (t \xx^{-1} + k-t) \, .
\label{EQ:DELTA}
\end{equation}
We also have
\begin{equation}
\V_z \gr{W}_{0,C_3}(z) = \frac{T^4}{2(1-T)^2} %
 = \frac{\xx^{-2}  }{2} - 2\xx^{-1} + 3 - 2 \xx + \frac{{\xx}^2}{2} \, .
\end{equation}
\begin{eqnarray}
& & (\V_z^2 - \V_z -2(k-1)) \xx^{-t} + %
2(\V_z \gr{W}_{0,C_3})(\V_z \xx^{-t}) = \cr
& & \, \, \, %
t(t+3)  \xx^{-t-4} - t(2t+8) \xx^{-t-3} + \cdots \, .
\end{eqnarray}
Using these formulae, the induction hypothesis, the form
of the generating function $6\gr{S}_{k,C_3} + 2\gr{J}_{k,C_3}$ and the
 formula (\ref{FUNCTIONAL_TRIANGLE_GRAPH}) of lemma \ref{LEMMA0},
when looking after the coefficients of 
$\xx^{-3k+1}$ and $\xx^{-3k}$, we find
\[
\gr{W}_{k,C_3} = b_k \xx^{-3k} - \cpt{k}  \xx^{-3k+1} + \cdots
\]
where the sequences $(b_k)$ and $(\cpt{k})$ satisfy 
exactly the recurrences given by (\ref{EQ:B_K}) and 
\begin{equation}
\begin{array}\{{rl}.
           &  \cpt{1}=\frac{25}{24} \, , \\
           &  \, \, %
2(3k+2)\cpt{k+1} = 8(k+1)b_{k+1} \\
& \, \, + \, \, 6 kb_k  + (3k-1)(3k+2)\cpt{k}  \\
& \, \, + \, \, 6 \sum_{t=1}^{k-1}t(3k-3t-1)b_t\cpt{k-t} \, .
\end{array}
\label{PRE_CPRIME_0}
\end{equation}
Now, we can show (\ref{CPRIME_0}) by induction. We have
$\cpt{1} = \frac{25}{24}$, $b_1 = \frac{5}{24}$ and
$c_2 = \frac{65}{48}$ and we can check 
$\cpt{2} = \frac{5}{3} = c_2 + \frac{3}{2} b_1$.
Suppose that for $i$ from $1$ to $k-1$, 
$\cpt{i}$ verifies
\[
\cpt{i+1} = c_{i+1} + \frac{3}{2} i b_i \, .
\]
Using (\ref{PRE_CPRIME_0}) and the induction hypothesis, we have
for $i=k$ (we have to be careful with $\cpt{1}=c_{1} + \frac{1}{4}$)
\begin{eqnarray}
2(3k+2)\cpt{k+1} = & & 8(k+1)b_{k+1} + 6kb_k \cr
 & &  + (3k-1)(3k+2)c_{k} + \frac{3}{2} (3k-1)(3k+2)(k-1)b_{k-1} \cr
 & &  + 12(k-1)b_{k-1}c_1 + 3(k-1)b_{k-1} \cr
 & &  + 6 \sum_{t=1}^{k-2} t(3k-3t-1) b_t c_{k-t} \cr
 & &  + 9 \sum_{t=1}^{k-2} t(3k-3t-1)(k-t-1) b_t b_{k-t-1} \, .
\end{eqnarray}
And as already remarked by Wright,
\cite[eq. (3.5)]{Wr80}, for any given sequence $(\alpha_k)$ we have
\begin{equation}
\sum_{t=1 }^{k-1 } t \alpha_t \alpha_{k-t} = %
\frac{ k}{2} \sum_{t=1 }^{k-1 }  \alpha_t \alpha_{k-t} \, .
\label{REMARK_ALPHA}
\end{equation}
Rearranging, we find using the definition of $c_{k+1}$ given by
(\ref{EQ:C_K}) and (\ref{REMARK_ALPHA})
\begin{eqnarray}
2(3k+2)\cpt{k+1} & & = 2(3k+2)c_{k+1} + 3kb_k  \cr
 & & + (3 + \frac{3}{2}(3k-1)(3k+2) )(k-1)b_{k-1} \cr
 & & + \frac{9}{2}(3k+1) \sum_{t=1}^{k-2} t b_t (k-t-1) b_{k-t-1} \, .
\end{eqnarray}
Since $3 \sum_{t=1}^{k-2} tb_t (k-t-1)b_{k-t-1} = 
2k b_k - 3(k-1)kb_{k-1}$, we obtain
\begin{eqnarray}
2(3k+2)\cpt{k+1} & & = 2(3k+2)c_{k+1} + 3kb_k  \cr
 & & + (3 + \frac{3}{2}(3k-1)(3k+2) )(k-1)b_{k-1} \cr
 & & + 3(3k+1)kb_k - \frac{9}{2}(k-1)k(3k+1) b_{k-1} \, .
\end{eqnarray}
Finally, we find $2(3k+2)\cpt{k+1} = 2(3k+2)c_{k+1} + 3(3k+2)kb_k$.  \ENDPROOF

As a consequence, if we want to work with a 
forbidden subgraph $H$ which is not unicyclic (e.g. $K_4$),
the decomposition of $\gr{W}_{k,H}$ into sums of
negative powers of $\xx$ (i.e. tree polynomials)
starts
\[
\gr{W}_{k,H} = b_k \xx^{-3k} - c_{k}  \xx^{-3k+1} + \cdots \, .
\]
The same remark holds for any finite collection
of forbidden subgraphs which are not unicyclic.

In the next theorem, we will generalize the case $\xi = \{ C_3 \}$.
\begin{thm} \label{THM_GENERAL_FORM_MANY}
Let $\xi = \{H_1,H_2,\cdots, H_p\}$ a finite collection of multicyclic 
components. Suppose that $\xi$ contains $r$, $r > 0$,
 distinct polygons (unicyclic smooth graphs).
Denote by $\gr{W}_{k,\xi}$ the EGF of $(k+1)$-cyclic
$\xi$-free labelled graphs. For all $k \geq 2$, $\gr{W}_{k,\xi}$ can
be expressed as a finite sum of powers of $\frac{1}{1-T}$ 
and has the following form:
For $k=1$, we have
\begin{equation}
\gr{W}_{1,\xi}(z) = \frac{5}{24 \big( 1-T(z)\big)^3} 
- \frac{\big(19/24 + r/4\big)}{\big(1-T(z)\big)^2}   
        + \sum_{i \leq 1} %
  \frac{{\psi_{i,1}}^{(\xi)}}{\big(1-T(z)\big)^{i}   }  \,  
\end{equation}
and for $k > 1$
\begin{equation}
\gr{W}_{k,\xi}(z) = \frac{b_k}{\big(1-T(z)\big)^{3k} } %
- \frac{ \cpxi{k}  }{ \big(1-T(z) \big)^{3k-1}    }        %
        + \sum_{i \leq 3k-2} %
\frac{{\psi_{i,k}}^{(\xi)}}{\big(1-T(z)\big)^{i}}  \, 
\label{FORME_GENERALE}
\end{equation}
where $b_k$ is Wright's coefficient
of first order given by (\ref{EQ:B_K}) and 
$\cpxi{k}$ is given recursively by $\cpxi{1} = \frac{19+6r}{24}$
and for $k \geq 1$
\begin{equation}
\cpxi{k+1} = c_{k+1} + \frac{3}{2} r k b_{k} \, .
\label{EQ:CPXI}
\end{equation}
\end{thm}

\noindent \textbf{Proof.} The proof of this theorem is very close
to that of theorem \ref{THM_GENERAL_FORM}. 
Suppose that $\xi$ contains $r$ polygons $(r > 0)$. Furthermore,
suppose that $C_q$ is the greatest polygon of $\xi$. That is
\[
\gr{W}_{0,\xi} = \frac{1}{2} \ln{\frac{1}{1-T}} - \frac{T}{2} %
- \frac{T^2}{4} - \sum_{i} \frac{T^i}{2i}
\]
where in the summation $i$ describes all lengths (less than 
or equal to $q$) of the forbidden polygons.
Then, since
\[
\frac{T^q}{(1-T)^2 } = \xx^{-2} - (q+1) \xx^{-1} +%
\sum_{j=1}^{q} T^{q-j} \, ,
\]
we have
\[
 2 \V_z \gr{W}_{0,\xi}(z) =%
\frac{T^{q+1} }{(1-T)^2} + \sum_{j} \frac{T^j}{1-T} \, 
\]
where the summation is over all lengths of the $q-r-2$ 
authorized (distinct) polygons. So,
\begin{equation}
  2 \V_z \gr{W}_{0,\xi}(z) = \xx^{-2} - (r+3)\xx^{-1} %
+ \mbox{Polynomial}_{\xi}(T) \, 
\end{equation}
and  $2 (\V_z \gr{W}_{0,\xi}(z))(\V_z \xx^{-t})$ starts with
\begin{equation}
t \xx^{-t-4} - (r+4) t \xx^{-t-3} + \cdots \, .
\end{equation}
Defining the operator $\Omegaxi{k}$ as
\begin{equation}
\Omegaxi{k}: \, \, \Omegaxi{k}  =%
 \Big( \big( \V_z^2 - 3 \V_z - 2k\big) +%
 2 \big( \V_z \gr{W}_{0,\xi}(z) \big) \V_z \Big)  \, 
\label{OP:OMEGAXI_K}
\end{equation}
and $\Lambdaxi{k}$ as the formal power seriers
\begin{equation}
\Lambdaxi{k}: \,  \Lambdaxi{k}(z) = \sum_{t=1}^{k-1} %
\Big(\V_z \gr{W}_{t,\xi}(z) \Big) %
\Big(\V_z \gr{W}_{k-t,\xi}(z)\Big) \, ,
\label{EQ:LAMBDAXI_K}
\end{equation}
we can generalize (\ref{AVEC_OP_C3}) 
\begin{eqnarray}
\Delta_{k+1} \gr{W}_{k+1,\xi} + 2 \sum_{\mathcal{H} \in \xi} %
e(\mathcal{H}) \gr{S}_{k+1,\mathcal{H}} %
+ 2 \gr{J}_{k+1,\xi} = \cr
\Omegaxi{k} \gr{W}_{k,\xi} + \Lambdaxi{k} \, , \, \, (k \geq 1) \, .
\label{AVEC_OP} 
\end{eqnarray}
Then, we find 
\begin{equation}
\Omegaxi{k}\xx^{-t} = t(t+3) \xx^{-t-4} - %
 t(2t+r+7) \xx^{-t-3} + \cdots 
\end{equation}
As for theorem \ref{THM_GENERAL_FORM}, we find that
$\cpxi{k+1}$ satisfies $\cpxi{1}=c_1 + \frac{r}{4}$ and for
$k \geq 1$
\begin{eqnarray}
& & 2 (3k +2) \cpxi{k+1} = 8(k+1) b_{k+1} + 3k(r+1)b_{k} + \cr
& & \, \,\,\,\,\,\, %
(3k-1)(3k+2) \cpxi{k} + 6\sum_{t=1}^{k-1} t(3k-3t-1) b_t \cpxi{k-t} \, . 
\label{EQ:PRE_CPXI}
\end{eqnarray}
We can now argue as for the proof of theorem \ref{THM_GENERAL_FORM}
to verify that the sequence $(\cpxi{k})$ satisfies
(\ref{EQ:CPXI}).
\ENDPROOF

In the next section, we will determine the asymptotic number of 
triangle-free labelled components 
when the number of exceeding edges satisfies $k=o(n^{1/3})$.
 
\section{Asymptotic number of sparsely connected labelled triangle-free
components}

The methods we give are based on the fundamental work of 
Wright in \cite{Wr80} with some  ingredients from analytic combinatorics.

First of all, we will study the behavior of 
\[
 t_n(a\, n+\beta)= n! \, \coeff{z^n} \frac{1}{(1-T(z))^{a \, n+\beta}} \, 
\]
where $a \equiv a(n)$ tends to $0$ as $n \rightarrow \infty$ 
and $\beta$ is fixed. Then, we will show that if 
$\beta_1 < \beta_2$,  
$a \equiv a(n) \rightarrow 0$ as $n \rightarrow \infty$
but  $\frac{a \, n}{{\ln{n}}^3} \rightarrow \infty$,
then $\frac{t_n(a \, n+ \beta_1)}{t_n(a \, n+\beta_2)} \rightarrow 0$.

Next, we will give a general framework analogous to that
of Wright in \cite{Wr80}. More precisely, let $(b_k)$
and $(\cpt{k})$ be the coefficients given by (\ref{EQ:B_K})
and (\ref{CPRIME_0}).  
We will show that 
the coefficients of the EGFs $\gr{W}_{k,C_3}$ 
 satisfy the following inequalities
\begin{eqnarray}
 & &  n! \coeff{z^n} \gr{W}_{k,C_3}(z)  \leq  n! \coeff{z^n} %
\frac{b_k}{\big(1-T(z)\big)^{3k}} \, \, \, \, \, \, \, and \cr
 &  & n! \coeff{z^n} \left( \frac{b_k}{\big(1-T(z)\big)^{3k}} - %
\frac{\cpt{k}}{\big(1-T(z)\big)^{3k-1}} \right) %
  \leq  n! \coeff{z^n} \gr{W}_{k,C_3}(z)  \, 
\label{COEFF_INEGALITES}
\end{eqnarray}
which we shall call \textit{Wright's inequalities} for triangle-free graphs.
Thus, the inequalities in (\ref{COEFF_INEGALITES}) and the fact that 
$\frac{t_n(a \, n - 1)}{t_n(a \, n)} \rightarrow 0$
imply  that almost all connected components
with $n$ vertices and $n+o(n^{1/3})$ edges 
are $\xi$-free whenever $k=o(n^{1/3})$. Equivalently, we will show
that the number $c_{C_3}(n,n+k)$ of triangle-free $(k+1)$-cyclic
graphs is asymptotically the same as the number $c(n,n+k)$ of 
$(k+1)$-cyclic general graphs computed by Wright in \cite{Wr80}
(see \cite{BCM90} for the extension of Wright's asymptotic results).

\subsection{Saddle point method for tree polynomials}
\label{SEC:COL}
In \cite{KP89}, Knuth and Pittel studied  combinatorially
and analytically the polynomial $t_n(y)$ defined  as follows
\begin{equation}
t_n(y) = n! \coeff{z^n} \frac{1}{\big( 1-T(z)\big)^y}
\label{EQ:TREE_POLYNOMIAL}
\end{equation}
which they call \textit{tree polynomial}.
In fact, the authors of \cite{KP89} observed that
the analysis of these polynomials can also be used 
 to study random graphs.

The lemma below is an application of the saddle point method 
\cite{Bruijn,FS+} to study the asymptotic behavior of the
coefficients $n! \coeff{z^n} \big(1-T(z) \big)^{-m(n)}$ as $m, \, n$
tend to infinity but $m=o(n)$. 
\begin{lem} \label{TREE_POLYNOMIAL_INFINITY}
Let $a \equiv a(n)$ such that $a \rightarrow 0$
but $\frac{a \, n}{{\ln{n}}^3} \rightarrow \infty$, and
 $\beta$ a fixed number. Then, the tree polynomial $t_n(a\,n+\beta)$
defined in (\ref{EQ:TREE_POLYNOMIAL}) satisfies
\begin{equation}
t_n(a \, n+\beta) = \frac{n!}{2 \sqrt{\pi n}} %
\frac{\exp{ (n u_0) } (1-u_0)^{(1-\beta)}}{{u_0}^n (1-u_0)^{a\,n}}  %
\left(1+O\big(\sqrt{a}\big) + O\big(\frac{1}{\sqrt{a\,n}}\big)\right)
\label{EQ:TREE_POLYNOMIAL_INFINITY}
\end{equation}
where $u_0 = 1 + \frac{a}{2} - \sqrt{a(1+\frac{a}{4})}$. 
\end{lem}

\noindent \textbf{Proof.} Cauchy's integral formula gives
\begin{eqnarray}
t_n(a\,n+\beta) & = & n! \coeff{z^n} %
\frac{1}{\big(1-T(z) \big)^{a\,n+\beta}} \cr
          & = & \frac{n!}{2 \pi i} %
      \oint \frac{1}{\big(1-T(z) \big)^{a\,n+\beta}} \frac{dz}{z^{n+1}} \,
\label{COL_UN}
\end{eqnarray}
where we integrate around a small circle enclosing the origin
and whose radius is smaller than $1/e$ (since $1/e$ is
the radius of convergence of the formal power series
$T(z) = \sum_{n \geq 1} n^{(n-1)} \frac{z^n}{n!}$).
We make the substitution $u=T(z)$ and
get $dz = e^{-u}(1-u)du$. Thus,
\begin{equation}
t_n(a\,n+\beta) = \frac{n!}{2 \pi i} \oint \frac{e^{nu}\, du}%
{(1-u)^{a\,n+\beta-1} \, u^{n+1}  } \, .
\label{COL_DEUX}
\end{equation}
The power $\big(\exp{(u)}/(1-u)^a \big)^n$ suggests us to use the
 saddle point method. We will describe briefly this
method for our case and refer to
de Bruijn \cite[Chap. 5]{Bruijn}, 
Flajolet and Sedgewick \cite{FS+} or Bender \cite{Be74}
for more details on general asymptotic methods.

We set $h(u) = u - \ln(u) - a \ln(1-u)$.
Starting with (\ref{COL_DEUX}), we now have
\begin{equation}
t_n(a\,n+\beta) = \frac{n!}{2 \pi i} %
\oint (1-u)^{1-\beta} \exp(n h(u))\frac{du}{u} \, .
\label{COL_TROIS}
\end{equation}
Let $F(r,\theta)$ be the integrand of
\begin{eqnarray}
 & & \frac{1}{2 \pi r^n} \int_{-\pi}^{\pi} (1- re^{i \theta})^{1-\beta} %
\exp(n h(r e^{i \theta}) ) d\theta  \cr
     & = & \frac{1}{2 \pi r^n} \int_{-\pi}^{\pi} F(r,\theta) d\theta \, .
\end{eqnarray}
The saddle point method consists to remark that
$F(r,\theta)$ turns very quickly as $n \rightarrow \infty$ 
 such that the essential of the integral is captured
by only few values of $\theta$, say 
$\theta \in \coeff{-\theta_0, \, \theta_0}$ (with $\theta_0 \rightarrow 0$).
Then, we have to choose the radius $r$
in order to concentrate the main contribution
of the integral, viz. for $\theta \in  \coeff{-\theta_0, \, \theta_0}$,
 $|F(r,\theta)|$ represents the essential of the integral.
In other words, we have to find a vicinity
of $\theta=0$ where  $| F(r,\theta) |$ takes its maximum.
Hence, we investigate the roots of $h^{'}(u) = 0$ and we find
two saddle points, at $u_0 = 1 + \frac{a}{2} - \sqrt{a(1+\frac{a}{4})}$
and $u_1 = 1 + \frac{a}{2} + \sqrt{a(1+\frac{a}{4})}$. We notice that
$h^{''}(u) = \frac{1-2u + (1+a)u^2}{u^2(1-u)^2}$,
$h^{''}(u_0) = 2 + 3 \sqrt{a} + O(a)$ and
$h^{''}(u_1) = 2 - 3 \sqrt{a} + O(a)$.
The main point of the application of the saddle
point method here is that $h^{'}(u_0)=0$ and
$h^{''}(u_0) > 0$, hence $nh(u_0 \exp{(i\theta)})$
is approximately $nh(u_0) - n {u_0}^2 h^{''}(u_0) \frac{\theta^2}{2}$
in the vicinity of $\theta = 0$.
If we integrate (\ref{COL_TROIS}) around a circle 
passing vertically through $u=u_0$, we obtain:
\begin{equation}
t_n(a\,n+\beta) = \frac{n!}{2 \pi i} 
\int_{-\pi}^{\pi} %
(1-u_0 e^{i \theta})^{1-\beta} \exp( n h(u_0 e^{i\theta}) ) d \theta \, 
\label{COL_QUATRE}
\end{equation}
where
\begin{equation}
h(u_0 e^{i\theta}) = u_0 \cos \theta + i u_0 \sin \theta %
- \ln u_0 - i \theta - a \ln (1-u_0 e^{i \theta}) \, \, .
\end{equation}
Denote by $\EuFrak{Re}(z)$ the real part of $z$, we have
\begin{eqnarray}
f(\theta) & = & \EuFrak{Re}( h(u_0 e^{i \theta}))  \cr
  & = & u_0 \cos \theta - \ln u_0 - %
a \ln ( |1-u_0e^{i\theta}|) \cr
 & = & u_0 \cos \theta  - \ln u_0 - %
 a \ln u_0 - \frac{a}{2} %
\ln \big( 1+ \frac{1}{u_0^2} - \frac{2}{u_0} \cos \theta  \big) \, .
\end{eqnarray}
It comes
\begin{equation}
f^{'}(\theta) = %
\frac{d}{d \theta}  \EuFrak{Re}(h(u_0 e^{i\theta})) = %
- u_0 \sin \theta - %
\frac{\frac{a}{2}\big( \frac{2}{u_0} \sin \theta \big)^2}%
{\big(1 + \frac{1}{u_0^2} - \frac{2}{u_0} \cos \theta \big)}
\end{equation}
and $f^{'}(\theta) = 0$ if $\theta = 0$. Also, $f(\theta)$ is
a symmetric function of $\theta$ and in 
$\left[ -\pi, -\theta_0 \right] \cup \left[ \theta_0, \pi \right]$, for
a given $\theta_0$, $0  < \theta_0 < \pi$, it takes it maximum value for
$\theta = \theta_0$. 
Since $|\exp( h(u))| = \exp( \EuFrak{Re}(h(u)) )$,
when splitting
the integral in (\ref{COL_QUATRE}) into three parts, viz.
``$\int_{-\pi}^{-\theta_0} + \int_{-\theta_0}^{\theta_0}
 + \int_{\theta_0}^{\pi}$'', we know that it suffices to integrate
from $-\theta_0$ to $\theta_0$, for a convenient value of
$\theta_0$, because
the others can be bounded by the magnitude of the integrand
at $\theta_0$. In fact, we have
\begin{eqnarray}
h(u_0 e^{i\theta}) & & = h(u_0)+ \frac{{u_0}^2 %
  (e^{i\theta}-1)^2}{2!}h^{''}(u_0) %
             +\frac{{u_0}^3 (e^{i\theta}-1)^3}{3!}h^{(3)}(u_0) \cr
                   & &   +\frac{{u_0}^4 (e^{i\theta}-1)^4}{4!}h^{(4)}(u_0) %
                   + \sum_{p \geq 5} %
\frac{{u_0}^p (e^{i\theta}-1)^p}{p!}h^{(p)}(u_0)   %
\cr
                   & & = h(u_0) + \sum_{p\geq 2} \alpha_p (e^{i\theta} -1)^p %
                   \, ,
\label{ETOILE4}
\end{eqnarray}
where $\alpha_p = \frac{{u_0}^p}{p!} h^{(p)}(u_0)$. We compute
$ h^{(p)}(u_0) = (-1)^p (p-1)! \Big( \frac{1}{{u_0}^{p}} -
\frac{a}{{(1-u_0)}^{p}}\Big)$, for $p \geq 2$. Then, on first hand we obtain
\begin{eqnarray}
\alpha_p & & =  \frac{ (-1)^p}{p} \Big(1 - \frac{a {u_0}^p}%
                                  {(1-u_0)^p} \Big) \cr
         & & = \frac{ (-1)^p}{p} + %
         \frac{ (-1)^{p+1}}{p} \, %
 \frac{a (1+\frac{a}{2} - \sqrt{a(1+\frac{a}{4})})^p}{{a}^{\frac{p}{2}} %
 (\sqrt{1+\frac{a}{4}} - \frac{\sqrt{a}}{2} )^p }  \cr
         & & =  \frac{(-1)^p}{p} + %
    \frac{(-1)^{p+1}}{p}  \, \frac{2^p}{{a}^{\frac{p}{2}-1}} \, %
\frac{(1+\frac{a}{2} %
- \sqrt{a(1+\frac{a}{4})})^p}{(\sqrt{1+\frac{a}{4}} - \frac{\sqrt{a}}{2} )^p } \, .
\label{ALPHA_P}
\end{eqnarray}
Hence, 
\begin{equation}
| \alpha_p | \leq O\Big( \frac{2^p}{a^{\frac{p}{2}-1}}\Big) \,,  %
\, \, \, (a \rightarrow 0) \,.
\end{equation}
On the other hand,
\begin{equation}
| e^{i\theta} - 1 | = \sqrt{ 2(1- \cos \theta)} < \theta %
\,, \,\, (\theta > 0)\, .
\end{equation}
Thus, the summation in (\ref{ETOILE4}) can be bounded for values of
$\theta$ and $a$ such that $\theta \rightarrow 0$, 
$a \rightarrow 0$ but $\frac{\theta}{\sqrt{a}} \rightarrow 0$ and we have
\begin{eqnarray}
| \sum_{p \geq 4} \alpha_p (e^{i\theta} - 1)^p | %
  & &    \leq \sum_{p \geq 4} | \alpha_p \theta^p | \cr
  & &    \leq  \sum_{p \geq 4} O\Big( \frac{2^p \theta^p}{a^{\frac{p}{2}-1}}\Big) %
       =  O\Big( \frac{\theta^4}{a} \Big) \, .
\end{eqnarray}
It follows that for $\theta \rightarrow 0$,
$a \rightarrow 0$ and $\frac{\theta}{\sqrt{a}} \rightarrow 0$
\begin{eqnarray}
h(u_0e^{i\theta}) & & = h(u_0) %
 - \frac{1}{2} \, \frac{u_0}{(1-u_0)^2} (1+a -2u_0+{u_0}^2) \theta^2 \cr
& & + i \frac{u_0}{6(1-u_0)^3} \, (1+a+(a-3)u_0+3{u_0}^2 - {u_0}^3) \theta^3 %
+ O\Big(\frac{\theta^4}{a}\Big)\, ,
\end{eqnarray}
where the term in the big-oh takes into account the terms from
$(e^{i\theta}-1)^2$ and $(e^{i\theta}-1)^3$ of (\ref{ETOILE4})
which we can neglect since $(e^{i\theta} -1)^2 = - \theta^2 - i \theta^3
+ O(\theta^4)$ and $(e^{i\theta}-1)^3 = -i\theta^3 + \frac{3}{2}\theta^4 +
iO(\theta^5)$.
Therefore, if $a \rightarrow 0$ but 
$\frac{an}{{(\ln{n})}^2} \rightarrow \infty$, if we let
$\theta_0 = \frac{\ln n}{\sqrt{n \rho}}$ with
$\rho = \frac{u_0(1+a-2u_0+{u_0}^2)}{(1-u_0)^2} = 2 - \sqrt{a} + O(a)$,
we can remark (as already said) that  it suffices to 
integrate (\ref{COL_QUATRE}) from $- \theta_0$ to
$\theta_0$, using the magnitude of the integrand at $\theta_0$
to bound the resulting error. Hence,
\begin{eqnarray}
& & |(1-u_0 e^{i\theta_0})^{(1-\beta)} %
 \left(\exp{( n h(u_0 e^{i \theta_0} ))}  - n u_0 + n \ln u_0 %
+ a \ln(1-u_0)  \right) |  =   \cr
& &   |1-u_0 e^{i\theta_0} |^{(1-\beta)}  %
\exp \Big( - \frac{n}{2} \rho \, {\theta_0}^2 + %
+ O\big( n \frac{{\theta_0}^4}{a} \big) \Big) = 
 O\Big( e^{-\frac{(\ln{n})^2}{2} } \Big)  \, .
\end{eqnarray}
To estimate $t_n(a\,n+\beta)$, it proves convenient to compute
\begin{equation}
J_n = \int_{-\theta_0}^{\theta_0} (1-u_0 e^{i \theta})^{(1-\beta)} %
\exp{(nh(u_0 e^{i\theta}))} d\theta \, .
\end{equation}
If we make the substitution $\theta = \frac{t}{\sqrt{n \rho}}$,
we have (recall that $\theta_0 = \frac{\ln n}{\sqrt{n \rho}} $)
\begin{equation}
J_n = \frac{1}{\sqrt{n \rho}} \int_{ - \ln{n}}^{\ln{n}} %
\Big( 1 - u_0  e^{\frac{it}{\sqrt{n\rho}}}\Big)^{(1-\beta)} %
\exp\Big(nh(u_0 e^{\frac{it}{\sqrt{n\rho}}}) \Big) dt \, .
\end{equation}
Since $(1-u_0 e^{\frac{it}{\sqrt{n\rho}}})^{(1-\beta)} = 
(1-u_0)^{(1-\beta)}(1+O(t/\sqrt{na})) $, $J_n$ becomes
\[
J_n  =  \frac{1}{\sqrt{n \rho}} \lambda_n
\]
where $\lambda_n = \int_{-\ln{n} }^{ \ln{n} }
(1-u_0)^{(1-\beta)} \exp\Big(nh(u_0) -\frac{t^2}{2} + 
 i f_3 \frac{t^3}{\sqrt{na}} + 
O\big(\frac{t^4}{na}\big) \Big) \Big(1+O\big(\frac{t}{\sqrt{na}}\big) \Big) \,
dt$ and 
$f_3 = - \frac{\sqrt{a}(1+a+(a-3)u_0+3u_0^2-u_0^3)}%
{\sqrt{u_0}(1+a-2u_0+u_0^2)^{\frac{3}{2}}}= %
- \frac{\sqrt{2}}{12} -\frac{5}{48}\sqrt{a} + O(a)$.
We obtain 
\begin{eqnarray}
J_n & = & \frac{(1-u_0)^{(1-\beta)}}{\sqrt{n\rho}} e^{(nh(u_0))} \, \, \times \cr
& & \left[ \, \int_{-\ln{n}}^{\ln{n} } e^{- \frac{t^2}{2}}%
 \cos{\big( f_3\frac{t^3}{\sqrt{na}}\big)}%
 \Big( 1+ O\big(\frac{t}{\sqrt{na}} \big) + %
O\big( \frac{t^4}{na}\big)\Big)\, dt \right] \cr
&=& \frac{(1-u_0)^{(1-\beta)}}{\sqrt{n\rho}} e^{(nh(u_0))} \, \, \times \cr
& & \left[ \, \int_{-\infty}^{\infty} e^{- \frac{t^2}{2}}%
 \Big( 1+ O\big(\frac{t}{\sqrt{na}} \big) + %
O\big( \frac{t^6}{na} \big)    \Big)\, dt %
\, \, + \, \, O\big( e^{- \frac{ (\ln{n})^2}{2} } \big)  \right]\cr
&=& \frac{\sqrt{2\pi} (1-u_0)^{(1-\beta)} e^{(nh(u_0))} } {\sqrt{n\rho}} %
\Big( 1+ O\big(\frac{1}{\sqrt{na}}\big)\Big) \cr
&=& \sqrt{\frac{\pi}{n}}(1-u_0)^{(1-\beta)} e^{(nh(u_0))} %
\Big( 1+ O\big(\sqrt{a}\big) + O\big( \frac{1}{\sqrt{na}} \big) \Big) \, .
\label{COL_THE_END}
\end{eqnarray}
We used $\cos{(x)} = 1+O(x^2)$ and $\exp{(O(x))} = 1+O(x)$ when $x=O(1)$.
Since $t_n(a\,n+\beta) = \frac{n!}{2 \pi} J_n$, the proof of 
lemma \ref{TREE_POLYNOMIAL_INFINITY} is now complete.
\ENDPROOF

\subsection{Wright's inequalities}
\label{INEGALITES_WRIGHT}
In order to adapt the techniques of Wright to
our $\xi$-free components, we need to
bound  the \textit{perturbative terms}, i.e., the EGFs containing
the first apparitions of the forbidden configurations
$\gr{S}_{k,\xi}$ and $\gr{J}_{k,\xi}$.

\subsubsection{Upper bounds of $\gr{S}_{k,\xi}$ and $\gr{J}_{k,\xi}$}
To take control on these EGFs,  let us recall briefly
the \textit{shrinking-and-expanding} Bagaev's method \cite{BV98}:
In order to enumerate
graphs of a given type, an induced subgraph with special properties should
be chosen and shrunk to a marked vertex. Separately, we
have to calculate:
\begin{itemize}
\item  the number of the obtained graphs, rooted at a fixed vertex of degree $d$,
\item  the number of the shrunk subgraphs,
\item  the number of ways to reconstruct the initial graphs.
\end{itemize}
We note that this technique generalizes the 
methods of lemmas \ref{LEMMA7} and \ref{LEMMA8}.  
\begin{figure}
    \begin{center} \epsfig{file= 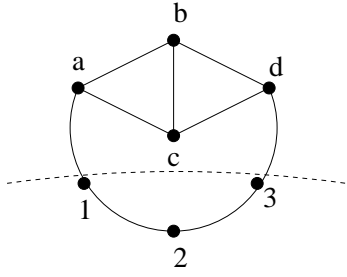,scale=0.50}
    \end{center}
    \caption[Illustration of Bagaev's method.]
    {Illustration of Bagaev's method.}
\label{FIG:BAGAEV}
\end{figure}
As an illustration of this method, consider the graph
depicted by figure \ref{FIG:BAGAEV} where $H$ is represented
by the juxtaposition of triangles. The number of ways to label
this graph can be computed easily using Bagaev's techniques. In fact,
we have
\[
{7 \choose 3} \times \underbrace{2 \times 1}_{reconstruction} %
\times 3 \times 6 = \frac{7!}{4} \, 
\]
manners to label the graph of  figure \ref{FIG:BAGAEV} (3 manners to label the
path with $3$ vertices and $6$ manners to label the juxtaposition of triangles).
This method is very useful to bound graph typified by the one in  
figure \ref{FIG:BAGAEV} (where our interest is focused on the juxtaposition
of triangles). The difficulties arise mainly from the number of 
possible reconstructions. In the current example, we have to rely
the vertices $1$ and $3$  to $2$ vertices belonging to 
$\{a, \, b, \, c, \, d\}$. Thus, the number of reconstructions is at most $4^2$
(including graphs different from the one in figure  \ref{FIG:BAGAEV}).

Consider now $\gr{S}_{k,\xi}$ with the special case
$\xi = \{C_3\}$.

\begin{lem}
\label{LEMMA_BOUND_S3}
For all $k \geq 1$ and $\forall \varepsilon > 0$
\begin{equation}
 \gr{S}_{k+1,C_3}  %
 \preceq   \Big(\frac{3}{2}+ \varepsilon\Big) \frac{kb_k}{{\xx}^{3k+2}} \, .
\label{EQ:LEMMA_BOUND_S3}
\end{equation}
\end{lem}

\noindent \textbf{Proof.} 
The bound of (\ref{EQ:LEMMA_BOUND_S3}) is inspired by the forms of
the EGF $\gr{S}_{k+1,C_3}$. We will prove (\ref{EQ:LEMMA_BOUND_S3})
by induction. We can verify that $\gr{S}_{2,C_3} \preceq \frac{5}{12\xx^5}$,
using  (\ref{eqn:SoloC3-2}). Suppose that
$\gr{S}_{i,C_3} \preceq \frac{2(i-1)b_{i-1}}{\xx^{3i-1}}$, for
$i \in \left[2,\, k-1\right]$ and let us prove that
$\gr{S}_{k,C_3}  \preceq   \frac{2(k-1)b_{k-1}}{{\xx}^{3k-1}}$.

Split the set
of $(k+1)$-cyclic graphs with exactly one occurrence of triangle
into three subsets as follows~:
\begin{itemize}
\item[1-] the first subset $\Sigma_1$
 contains all graphs whose situations
after smoothing are characterized by the fact that
 exactly one vertex of the triangle  is of degree $ \geq 3$,
\item[2-] similarly, the second subset $\Sigma_2$
is built with all graphs whose situations
after smoothing are characterized by the fact that
 exactly two vertices of the triangle are of degree $ \geq 3$,
\item[3-] $\Sigma_3$ contains all other graphs of $\mathcal{\gr{S}}_{k,C_3}$
not in $\Sigma_1 \cup \Sigma_2$.
\end{itemize}
We can bound the number of the graphs of the subsets $\Sigma_1$ and 
$\Sigma_2$,  using lemmas \ref{LEMMA7}, \ref{LEMMA8},
$\sth{\gr{W}_{k-1,C_3}} \preceq \frac{b_{k-1}}{ {(1-z)}^{3k-3} }$ 
(since Wright showed $\gr{W}_{k-1} \preceq \frac{b_{k-1}}{{\xx}^{3k-3}}$
\cite{Wr80}) 
and the fact that
$\V_z (\frac{1}{ {\xx}^t}) \preceq \frac{t}{{\xx}^{t+2}}$ for $t\geq 0$. In fact,
\begin{eqnarray}
\Sigma_1(z) + \Sigma_2(z) & & =  {\left[ \frac{1}{z(1-z)} \left(\V_z %
 \frac{b_{k-1}}{(1-z)^{3k-3}} \right) \,  \left( \V_z %
 \frac{z^3}{3!}\right) \right]}_{| z=T} + %
 \cr
& &    %
{\left[ \frac{2}{wz^{2}} \, %
\left( \V_w \frac{w^{k-1} b_{k-1}}{(1-wz)^{3k-3}}\right)%
 \, %
\left( \V_w %
 \frac{w^3 z^3}{3!}\right)\right]}_{| wz=T} \cr
& & \, \, \,  %
\preceq \frac{3}{2} \, \frac{(k-1)b_{k-1}}{ {\xx}^{3k-3}} \big( {\xx}^{-2} -
{\xx}^{-1} + \frac{5}{3} - \xx \big) \cr
& &  \preceq \frac{3}{2} \, \frac{(k-1)b_{k-1}}{ {\xx}^{3k-1}} \, .
\label{EQ:SIGMA1SIGMA2}
\end{eqnarray}
For graphs of $\Sigma_3$, we have two subcases. Denote by
$\Sigma_{3}^{'}$, resp. $\Sigma_{3}^{''}$,  the graphs 
of $\Sigma_3$ such that the deletion
of the $3$ vertices and the $3$ edges of the triangle will
leave a connected graph, resp. disconnected graphs.
The figures \ref{FIG:solo2C3} (c) and
\ref{FIG:solo2C3} (e) illustrate these 2 classifications.
In the first case, i.e. $\Sigma_{3}^{'}$, we will not use
the induction hypothesis. In fact, to build a graph of
 $\Sigma_{3}^{'}$, we have to rely $d$ vertices ($d \geq 3$) of
a graph of $\mathcal{\gr{W}}_{k-d,C_3}$ to the triangle.
Thus, the number of manners to construct a graph of  $\Sigma_{3}^{'}$
of order $n$ this way is at most
\begin{eqnarray}
& & 3^d {n \choose 3} {{n-3} \choose d} (n-3)! %
 \coeff{z^{n-3}}\gr{W}_{k-d,C_3}(z) \cr 
& & \leq \frac{3^d}{6} {{n-3} \choose d} n! %
 \coeff{z^{n-3}}\gr{W}_{k-d,C_3}(z) \cr 
& & \leq  \frac{3^d}{6 \, d!} n^d \, n! %
\coeff{z^{n}}\gr{W}_{k-d,C_3}(z) \cr
& & \leq \frac{3^d}{6 \,d!}  n! %
\coeff{z^{n}} \V_z^d \, \gr{W}_{k-d,C_3}(z) \, , (3 \leq d \leq k+1) \, .
\end{eqnarray}
In terms of generating function, we then have (summing over $d$)
\begin{eqnarray}
\Sigma_{3}^{'}(z) & & \preceq \sum_{d \geq 3} \frac{3^d}{6 \, d!} %
 \V_z^d \, \gr{W}_{k-d,C_3}(z) \, .
\label{SUM_OVER_D}
\end{eqnarray}
First, let us treat the cases $d=k+1$ and $d=k$.
We have
\[
\V_z^{(k+1)} \, \gr{W}_{-1} = \V_z^{k} \, T =  \V_z^{k-1}\, \frac{T}{\xx} \, 
\]
and
\[
 \V_z^{k} \, \gr{W}_{0,C_3} = \V_z^{k-1}\, \Big( \frac{T^4}{2 \xx^2 }\Big) \, .
\]
Since $\frac{T}{X} \preceq \frac{1}{\xx^2}$, we have
\[
\frac{3^k}{6 k!} \Big(1+\frac{3}{k+1} \Big)%
 \Big( \V_z^{(k+1)} \, \gr{W}_{-1} +  \V_z^{k} \, \gr{W}_{0,C_3}\Big) \preceq %
\frac{3^k}{k!} \V_z^{(k-1)} \Big( \frac{1}{\xx^2}\Big) \, .
\]
Similarly
\[
\V_z^{(k-1)} \frac{1}{\xx^2}  \preceq  \V_z^{(k-2)} \frac{2}{\xx^4} %
                              \cdots   %
     \preceq  \frac{2 \times 4 \times \cdots \times 2(k-1)}{\xx^{2k}} %
     = \frac{2^k(k-1)}{\xx^{2k}} \, 
\]
and we obtain for $d=k+1$ and $d=k$ in (\ref{SUM_OVER_D})
\begin{equation}
\frac{3^{k+1}}{6(k+1)!} \V_z^{(k+1)} \, \gr{W}_{-1} + 
\frac{3^{k}}{6k!} \V_z^{k} \, \gr{W}_{0,C_3} \preceq %
\frac{6^k}{6 k!} \frac{(k-1)!}{\xx^{2k}} \, .
\end{equation}
Next, we have
\[
\frac{b_{k+1}}{b_k} \geq \frac{3}{2}k 
\]
since  $b_k = (\frac{3}{2})^k (k-1)! d_k$ and $(d_k)$ is an
increasing sequence (cf. \cite[eq. (1.4)]{Wr80}). Thus,
\[
b_k  \geq  \frac{3}{2}(k-1) b_{k-1}  \geq  %
{(\frac{3}{2})}^2 (k-1)(k-2) b_{k-2} %
    \geq  \cdots  \geq  {(\frac{3}{2})}^{k-1} (k-1)! \, \, b_1
\]
and 
\begin{equation}
(k-1)! \leq 6 (k-1) b_{k-1} \, .
\label{FRAC_VS_B}
\end{equation}
Finally,
\begin{equation}
\frac{3^{k+1}}{6(k+1)!} \V_z^{(k+1)} \, \gr{W}_{-1} + 
\frac{3^{k}}{6k!} \V_z^{k} \, \gr{W}_{0,C_3} \preceq %
\frac{6^k}{6 k!} \frac{(k-1)b_{k-1}}{\xx^{2k}} \, .
\label{ETOILE-ETOILE}
\end{equation}
Summing (\ref{SUM_OVER_D}) over $d$ for $d \in \left[3, \, k-2\right]$,
we obtain
\begin{eqnarray}
& &  \sum_{d=3}^{k-2} \frac{3^d}{6 \, d!} %
 \V_z^d \, \gr{W}_{k-d,C_3}(z)
  \preceq  \sum_{d=3}^{k-2} \frac{3^d}{6 \, d!} %
 \V_z^d \, \frac{b_{k-d}}{\xx^{3k-3d}} \cr
 & & \preceq  \sum_{d=3}^{k-2} \frac{3^d}{6 \, d!} %
\frac{(3k -3d)(3k-3d+2) \cdots (3k-3d+2(d-1)) b_{k-d}}{\xx^{3k-d}} \cr
 & &  \preceq  \sum_{d=3}^{k-2} \frac{3^d}{6 \, d!} %
 \frac{3^d (k-d)(k-d+\frac{2}{3})(k-d+\frac{4}{3}) \cdots %
  (k - \frac{1}{3}d  -\frac{2}{3}) b_{k-d}}{\xx^{3k-d}} \cr
 & & \preceq \sum_{d=3}^{k-2} \frac{3^d}{6 \, d!} \, \, %
 3^d \frac{ (k-d)(k-d+1)(k-d+2) \cdots (k-1) b_{k-d}}{\xx^{3k-3}} 
\label{EQ:131}
\end{eqnarray}
So using (\ref{FRAC_VS_B}) and (\ref{ETOILE-ETOILE}), we get
after a bit of algebra
\begin{equation}
\Sigma_{3}^{'} \preceq %
\sum_{d=3}^{k+1} \frac{6^{d-1}}{d!} \, \, \frac{(k-1)b_{k-1}}{\xx^{3k-3}} %
\preceq 379 \, \frac{(k-1)b_{k-1}}{\xx^{3k-3}} \, .
\end{equation}
We can apply the same techniques as above for graphs of $\Sigma_{3}^{''}$.
However, we need here the help of the induction hypothesis where we will choose
$\varepsilon=\frac{1}{2}$ for sake of simplicity.
\begin{figure}
    \begin{center} \epsfig{file= 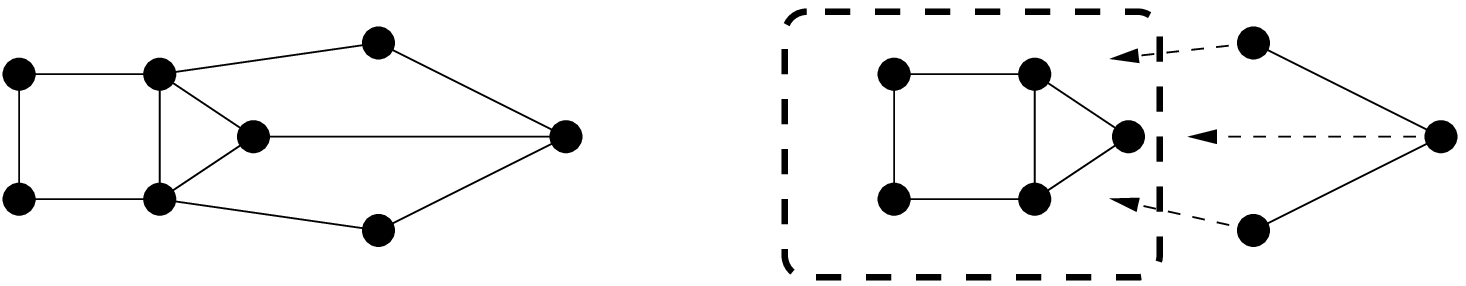,scale=0.50}
    \end{center}
    \caption[A representative graph of $\Sigma_{3}^{''}$ and its %
    reconstruction.]
    {A representative graph of $\Sigma_{3}^{''}$ and its %
    reconstruction.}
\label{FIG:SIGMA31}
\end{figure}  
In fact, a graph from  $\Sigma_{3}^{''}$ can be seen
as the composition of  two graphs: the first from $\mathcal{\gr{S}}_{e_1,C_3}$
and the second from $\mathcal{\gr{W}}_{e_2,C_3}$ (e.g. the graph
in the dashed box of figure \ref{FIG:SIGMA31}). 
Furthermore,  suppose that the first graph is of order
$p$, the second $n-p$ and that we have to rely $d$ vertices of the
second to the triangle (e.g. in the figure \ref{FIG:SIGMA31}, 
$d=3$, $p=5$ and $n=8$). 
The number of manners to label such composition is less than or equal to
\begin{eqnarray}
& & 3^d {n \choose p}{ {n-p} \choose d} %
p! \coeff{z^p} \gr{S}_{e_1,C_3} \, %
(n-p)! \coeff{z^{n-p}} \gr{W}_{e_2,C_3} \cr
& & \leq 3^d {  {n-p} \choose d} n! \coeff{z^n} %
\gr{S}_{e_1,C_3} \, \times \, \gr{W}_{e_2,C_3} \cr
& & \leq  \frac{3^d}{d!} \,  n! \coeff{z^n} \V_z^d \, %
(\gr{S}_{e_1,C_3} \, \times \, \gr{W}_{e_2,C_3}) \, .
\end{eqnarray}
We have $d+e_1+e_2=k$ and using the induction hypothesis on
$\gr{S}_{e_1,C_3}$ with the fact that 
$\gr{W}_{e_2,C_3} \preceq \gr{W}_{e_2} \preceq \frac{b_{e_2}}{\xx^{3e_2}}$, 
we obtain
\begin{eqnarray}
\Sigma_{3}^{''} %
& & \preceq \sum_{d+e_1+e_2=k} %
 \frac{3^d}{d!} \V_z^d (\gr{S}_{e_1,C_3} \, \times \, \gr{W}_{e_2,C_3}) \cr
& & \preceq  \sum_{d+e_1+e_2=k} %
 \frac{3^d}{d!} \V_z^d %
\frac{2(e_1-1)b_{e_1-1}b_{e_2}}{\xx^{3e_1+3e_2-1}} \cr
& &   \preceq 2 \, \sum_{d+e_1+e_2=k} %
  \frac{3^d}{d!}  (3k-3d-1)\cdots (3k-d-3)%
\frac{(e_1-1)b_{e_1-1}b_{e_2}}{\xx^{3k-d-1}} \cr
& & \preceq 2\, \sum_{d+e_1+e_2=k} %
 \frac{3^d}{d!} 3^d (k-d-\frac{1}{3})%
\cdots (k-\frac{d}{3}-1)%
\frac{(e_1-1)b_{e_1-1}b_{e_2}}{\xx^{3k-d-1}} \cr
& & \preceq 2\, \sum_{d+e_1+e_2=k} %
\frac{3^d}{d!} 3^d (k-d)(k-d+1) %
\cdots (k-1) \frac{b_{k-d}}{\xx^{3k-d-1}} \, ,
\end{eqnarray}
because we  have 
\begin{equation}
(e_1-1)b_{e_1-1}b_{e_2} \leq b_{k-d}
\end{equation}
since
\begin{eqnarray}
(e_1-1)b_{e_1-1} \, b_{e_2} & & = %
\Big( \frac{3}{2}\Big)^{e_1+e_2-1}%
(e_1-1)!\, (e_2-1)! \, d_{e_1-1} \, d_{e_2} \cr
 & & \leq %
\Big( \frac{3}{2}\Big)^{k-d-1} %
(e_1-1)! \, e_2!  \, d_{e_1-1} \, d_{e_2} \cr
 & & \leq %
\Big( \frac{3}{2}\Big)^{k-d-1} %
(e_1+e_2)! \, d_{e_1+e_2-1} \cr
 & & \leq \Big( \frac{3}{2}\Big)^{k-d} %
(e_1+e_2)! \, d_{k-d} \cr
 & & = b_{k-d} \, .
\end{eqnarray}
(We used $ (k+1)!d_{k+1} = (k+1)!d_k + %
\sum_{h=1}^{k-1} h! (k-h)! d_h d_{k-h}$ \cite[eq. (1.4)]{Wr80}.)
Hence,
\begin{equation}
\Sigma_{3}^{''} \preceq 2 \sum_{d \geq 1} \frac{6^d}{d!} \, %
 \frac{kb_k}{\xx^{3k-2}} %
\preceq 805 \frac{(k-1)b_{k-1}}{\xx^{3k-2}}\, .
\end{equation}
We have $ \coeff{z^n} \frac{1}{\xx^{3k-3}} %
\Big(\frac{\varepsilon}{{\xx}^2}-\frac{805}{\xx} - 379 \Big) \geq 0$,
$\forall n \geq 1$ since $\forall n > 0, \, %
\coeff{z^n}(\varepsilon-805 \xx - 379 \xx^2) \geq 0$ 
and 
$\coeff{z^n}\gr{S}_{k,C_3} = 0$ for $0 \leq n \leq 2$. (In fact,
$\forall a, \, b, \,c > 0$, we have 
$0 \preceq a-b\xx-c\xx^2 = (a-b-c)+bT+2c(T-\frac{T^2}{2}) \, .)$
Finally, we obtain
  $\gr{S}_{k,C_3}  \preceq   %
\Big(\frac{3}{2}+ \varepsilon\Big) \frac{(k-1)b_{k-1}}{{\xx}^{3k-1}}$. \ENDPROOF

\noindent
By similar methods, one can prove 
\begin{lem}
\label{LEMMA_BOUND_J3}
 For $\varepsilon > 0$ and $k\geq 2$,
\begin{equation}
\gr{J}_{k+1,C_3} \preceq \Big(6+\varepsilon\Big) %
\frac{(k-1)b_{k-1}}{\xx^{3k-1}} \, .
\label{EQ:LEMMA_BOUND_J3}
\end{equation}
\end{lem}
Before proving lemma \ref{LEMMA_BOUND_J3}, we notice that
working with juxtaposition of $t$ triangles as subgraph is much
easier. 

\begin{defn}
Denote by 
${\gr{J}}_{k,C_3}^{(t)}$ the EGF that counts $k$-excess graphs with
a juxtaposition of \textit{exactly} $t$ triangles sharing a common edge.
\end{defn}
(For instance,  the graph of figure \ref{FIG:BAGAEV} belongs to
the family ${\gr{\mathcal{J}}_{2,C_3}}^{(2)}$.)

\begin{lem}
\label{WITH_T_TRIANGLES}
$\forall \varepsilon > 0$, $k>t>1$, $k \geq 3$, we have
\begin{equation}
{\gr{J}_{k,C_3}}^{(t)} \preceq  (3+\varepsilon) %
\, \, \frac{(t+2)}{2!t!} \, \, \, %
 \frac{(k-t)b_{k-t}}{\xx^{3k-3t+2}}  \, .
\label{EQ:WITH_T_TRIANGLES}
\end{equation}
\end{lem}

\noindent \textbf{Proof (sketch).}  Smooth members of
${\gr{\mathcal{J}}_{t-1,C_3}}^{(t)}$ are counted by
\begin{equation}
\sth{\gr{\mathcal{J}}_{t-1,C_3}}^{(t)}(w,z) = %
\frac{w^{2t+1} z^{t+2}}{2! t!} \, .
\end{equation}
Thus, the reader can remark that the bound in 
(\ref{EQ:WITH_T_TRIANGLES}) is suggested by serial concatenation of
graphs of ${\gr{\mathcal{J}}_{t-1,C_3}}^{(t)}$ and of 
${\gr{\mathcal{W}}}_{k-t,C_3}$. At this stage, (\ref{EQ:WITH_T_TRIANGLES})
can be proved as it was be done for the bound of $\gr{S}_{k,C_3}$ in
lemma (\ref{LEMMA_BOUND_S3}). The main change is that the ``unique occurrence
of triangle'' has been replaced by a ``unique occurrence of
juxtaposition of $t$ triangles'' with EGF $\frac{w^{2t+1} z^{t+2}}{2! t!}$.
\ENDPROOF

\noindent \textbf{Proof of lemma \ref{LEMMA_BOUND_J3}.} 
It suffices to sum over all possible values of $t$. We have
\begin{eqnarray}
\gr{J}_{k,C_3} & \preceq & \frac{(3+\frac{\varepsilon}{2})}{2 \xx^{3k-4}} %
    \sum_{t=2}^{k} \frac{ (t+2)(k-t)b_{k-t}}{t!} \quad %
 (\mbox{we use  } \frac{1}{\xx^{3k-3t+2}} \preceq \frac{1}{\xx^{3k-4}} ) \, , %
 \cr
 & \preceq & \frac{(3+\frac{\varepsilon}{2}) (k-2)b_{k-2}}{\xx^{3k-4}} %
\sum_{t=2}^{k} \frac{t}{t!} \preceq %
\frac{(6+\varepsilon)(k-2)b_{k-2}}{\xx^{3k-4}} \, .
\end{eqnarray}
\ENDPROOF

Lemmas \ref{LEMMA_BOUND_S3} and \ref{LEMMA_BOUND_J3} suggest
themselves for generalization for any finite set $\xi$.
Although, we do not intend to present such generalization here,
we are convinced that this can be done practically in the same ways
as we did for (\ref{EQ:LEMMA_BOUND_S3}) and (\ref{EQ:LEMMA_BOUND_J3}).

\subsubsection{Bounds of $\gr{W}_{k,\xi}$}

In this paragraph, we present results that are strongly related
to those of Wright. In fact, the Wright's seminal paper
contains general techniques that are well suited for our triangle-free
graphs. In paragraph \S \ref{SUBSUB:GENERAL-FORM},
we obtained the general forms of the EGFs $\gr{W}_{k,\xi}$
(see theorem \ref{THM_GENERAL_FORM_MANY}).
Recall that $(b_k)$ and $(c_k)$ are given respectively by
(\ref{EQ:B_K}) and (\ref{EQ:C_K}). The lemmas
\ref{LEMMA_0} -- \ref{LEMMA_F} stated below 
will serve us to show by induction
the inequalities (\ref{COEFF_INEGALITES}). Before,
let us specify some useful notations.

\noindent \textbf{Notations.}


\noindent  For all $k \geq 1$, define by ${\EuScript{L}}_k$
and ${\EuScript{R}}_k$ the generating functions given by
(recall that $\xx = 1-T$)
\begin{equation}
{\EuScript{L}}_k(z) = \gr{W}_{k,C_3}(z) - \frac{b_k}{{\xx}^{3k}} %
+ \frac{\cpt{k}}{{\xx}^{3k-1}} \, 
\label{GAUCHE}
\end{equation}
and
\begin{equation}
{\EuScript{R}}_k(z) =  \frac{b_k}{{\xx}^{3k}}  - W_{k,C_3}(z) \, .
\label{DROITE}
\end{equation}

\noindent Recall that we just have to prove
that $\mathbf{{\EuScript{L}}_k \succeq 0}$ 
for all ${k \geq 1}$ since $\mathbf{{\EuScript{R}}_k \succeq 0}$ was proved by
Wright \cite{Wr80}. 
 
First of all, the following lemma gives bounds of $\cpxi{k}$ by means of $b_k$:
\begin{lem}
\label{LEMMA_0}
For all $k  \geq 1$, we have $kb_k \leq \cpxi{k} \leq \frac{19+6r}{5} kb_k$,
where $r$ is the number of polygons of $\xi$.
\end{lem}

\noindent \textbf{Proof.} 
We let $\cpxi{k} = k b_k (1+\betaxi{k})$. 
Hence, $\betaxi{1}= \frac{14+6r}{5}$
(where $r$ is the number of the forbidden polygons of distinct lengths).
After a bit of algebra, we find
\begin{eqnarray}
 2(3k+2)(k+1) b_{k+1}& & (1+ \betaxi{k+1})   =  8(k+1)b_{k+1} +3k(r+1)b_k \cr
    &   &  + (3k-1)(3k+2)kb_k (1+ \betaxi{k}) \cr
    &   &  + 6 \sum_{t=1 }^{k-1 } t(k-t)(3k-3t-1) b_t b_{k-t} (1 + \betaxi{k-t} )\, .
\end{eqnarray}
Let $\mathcal{B}_k$ and $\Ckxi{k}$ be
the rational numbers defined with the help of $(b_k)$ and
$(\cpxi{k})$ by
\begin{equation}
\mathcal{B}_k = \sum_{t=1}^{k-1} t (k-t) b_t b_{k-t} \, .
\label{BB_K}
\end{equation}
\begin{equation}
\Ckxi{k} = \sum_{t=1}^{k-1} t (3k - 3t -1) b_t \cpxi{k-t} \, .
\label{CC_K}
\end{equation}
Using (\ref{REMARK_ALPHA}), we find
\begin{eqnarray}
6  \sum_{t=1}^{k-1} & & t (k-t) (3k - 3t -1) b_t b_{k-t} %
 =  3(3k-2) \mathcal{B}_k \cr
& & =  2(k+1)(3k-2) b_{k+1} - 3k(k+1)(3k-2) b_k \, .
\end{eqnarray}
Thus, 
\begin{eqnarray}
& & 2 (3k+2)(k+1)b_{k+1} \betaxi{k+1}  =   %
(3r+7)kb_k + k(3k+2)(3k-1)b_k \betaxi{k} \cr
         &  &  + 6 \sum_{t=1}^{k-1} %
 t (k-t) (3k - 3t -1) b_t b_{k-t} \betaxi{k-t}
\label{eq:**}
\end{eqnarray}
and we have $\betaxi{k} > 0 $ for all $\xi$ and $k > 0$.
We let $\Rhoxi{k} = \mbox{max}_{1 \leq t \leq k} %
\betaxi{t} \geq \frac{14+6r}{5}$.
Then, (\ref{eq:**}),  (\ref{REMARK_ALPHA}) and (\ref{EQ:B_K}) give
\begin{eqnarray}
 2 (3k+2)(k+1)b_{k+1} \betaxi{k+1}  %
  & \leq &   (3r+7)kb_k + \cr
  & & \Big( k(3k+2)(3k-1)b_k \cr
  & & + 6 \sum_{t=1}^{k-1} t (k-t) (3k - 3t -1) b_t b_{k-t} \Big) %
  \Rhoxi{k}  \cr
  & \leq &   (3r+7)kb_k + \cr
  & &\Big( k(3k+2)(3k-1)b_k + 3(3k-2) \mathcal{B}_k\Big)  \Rhoxi{k}  \cr
  &  \leq  &   (3r+7)kb_k \cr 
  & +&  (2(3k-2)(k+1)b_{k+1} + 4kb_k) \Rhoxi{k}  \, .
\end{eqnarray}
Now, if we suppose that $\betaxi{k+1} > \Rhoxi{k}$, we will have
\begin{eqnarray}
8(k+1)b_{k+1}\betaxi{k+1} & \leq & 4k b_k \betaxi{k+1} + (3r+7)kb_k  \cr
12k(k+1)b_{k}\betaxi{k+1} + 12 \mathcal{B}_k \betaxi{k+1} %
& \leq & 4k b_k \betaxi{k+1} + (3r+7)kb_k  \, 
\end{eqnarray}
so that
\begin{equation}
12(k+1)b_{k}\betaxi{k+1} \leq  4 b_k \betaxi{k+1} + (3r+7)b_k  \, 
\end{equation}
and
\begin{equation}
\betaxi{k+1} \leq \frac{3r+7}{4(3k+2)}
\end{equation}
which is in contradiction with the fact that 
$\betaxi{k+1} > \Rhoxi{k} \geq \frac{14+6r}{5}$
(this will lead us to $3r+7 > 18kr + 42k$).
So, $(\betaxi{k})$ is a nonincreasing 
sequence and $\Rhoxi{k} = \frac{14+6r}{5}$ for all $k > 0$.
\ENDPROOF

Next, we have the lemmas
\ref{LEMMA_A}, \ref{LEMMA_B}, \ref{LEMMA_C}, \ref{LEMMA_D}, \ref{LEMMA_E} 
 stated  below, corresponding to the lemmas 6, 7, 8, 9 and 10 of
\cite{Wr80} but adapted for our $\xi$-free graphs. Lemmas 3 
and 4 of \cite{Wr80} are contained in lemma \ref{LEMMA_F}.

\begin{lem} 
\label{LEMMA_A}
If $\Big( \gr{W}_{t,\xi} - %
\frac{b_t}{{\xx}^{3t}} %
+ \frac{ \cpxi{t} }{{\xx}^{3t-1}} \Big) %
\succeq 0$, for $t$ such that $1 \leq t \leq k-1$ then
\begin{equation}
\Lambdaxi{k} \succeq \frac{T^2}{{\xx}^{3k+4}} %
 \, \Big(9 \mathcal{B}_k - 6 \Ckxi{k} \, {\xx} \Big) \, 
\label{LEMMA6_WRIGHT}
\end{equation}
where $\Ckxi{k}$ is given by (\ref{CC_K}) and $\Lambdaxi{k}$
is given by (\ref{EQ:LAMBDAXI_K}).
\end{lem}

\noindent \textbf{Proof.} If $x_1, \, \cdots, \, x_6$ are
positive real numbers and $x_1 \geq x_2 -x_3$, $x_4 \geq x_5 - x_6$ then
\begin{equation}
x_1 x_4 \geq x_2 x_5 - x_2 x_6 - x_5 x_3 \, .
\label{42}
\end{equation}
In fact, if $x_2 < x_3$ and/or $x_5 < x_6$, the right side of
the above inequality is negative. Otherwise,
if $x_2 \geq x_3$ and $x_5 \geq x_6$, we have:
\[
x_1 x_4 \geq (x_2 - x_3)(x_5 - x_6) \geq x_2 x_5 - x_2 x_6 - x_5 x_3 \, .
\]
Assume now that $1 \leq t \leq k-1$. We have
$\gr{W}_{t,\xi} \succeq 0$, $b_t/ {\xx}^{3t} \succeq 0$,
$(c_t + 3/2r (t-1)b_{t-1})/{\xx}^{3t-1} \succeq 0$ and
${\EuScript{L}}_t \succeq 0$ for $1 \leq t \leq k-1$.
Consequently, the coefficients of $\V_z \gr{W}_{t,\xi}(z)$ are
positive for the same value of $t$.
Setting
\begin{eqnarray}
x_1 & =&  s! \coeff{z^s} \V_z \gr{W}_{t,\xi}(z) \, ,  \cr
x_2 &=& b_t s! \coeff{z^s} \V_z \frac{1}{{\xx}^{3t}} \, , \cr
x_3 &=& \cpxi{t} s! \coeff{z^s} %
\V_z \frac{1}{{\xx}^{3t-1}} \, , \cr
x_4 &=& (n-s)! \coeff{z^{n-s}} \V_z \gr{W}_{k-t,\xi}(z) \, , \cr
x_5 &=& b_{k-t} (n-s)! \coeff{z^{n-s}} \V_z \frac{1}{{\xx}^{3k-3t}} %
\mbox{ and } \cr
x_6 &=& \cpxi{k-t} %
 (n-s)! \coeff{z^{n-s}} \V_z \frac{1}{{\xx}^{3k-3t-1}} \, 
\end{eqnarray}
where $s \in \coeff{0, \, n}$, after  substituting the values of $x_i$,
$i \in \coeff{1, \, 6}$ in (\ref{42}) and summing over  $s$
and $t$, $t \in \coeff{1, \, k-1}$, we obtain (\ref{LEMMA6_WRIGHT}).
 \ENDPROOF

\noindent Similarly, we have
\begin{lem} 
\label{LEMMA_B}
If $\Big(\frac{b_t}{{\xx}^{3t}} - \gr{W}_{t,\xi} \Big) \succeq 0$
for $1 \leq t \leq k-1$ then
\begin{equation}
\Lambdaxi{k} \preceq 9 \mathcal{B}_k \frac{T^2}{{\xx}^{3k+4}} \, .
\label{LEMMA7_WRIGHT}
\end{equation}
\end{lem}
In the following lemmas, we work again with the
special case $\xi=\{C_3\}$ for sake of clarity.
\begin{lem} 
\label{LEMMA_C}
Define by $\Yt{k}$ and $\Zt{k}$ the formal power series
\begin{eqnarray}
\Yt{k}(z) & &  = \Delta_{k+1} \frac{b_{k+1}}{{\xx}^{3k+3}} %
- \Omegat{k} \frac{b_k}{{\xx}^{3k}} %
 - 9 \mathcal{B}_k \frac{T^2}{{\xx}^{3k+4}}
\label{LEMMA8Y_WRIGHT}
\end{eqnarray}
\begin{eqnarray}
\Zt{k}(z) & & = \Delta_{k+1} \frac{\cpt{k+1}}{{\xx}^{3k+2}} %
- \Omegat{k} \frac{\cpt{k}}{{\xx}^{3k-1}} %
- 6\Ckt{k} \frac{T^2}{{\xx}^{3k+3}} \, .
\label{LEMMA8Z_WRIGHT}
\end{eqnarray}
For all $k \geq 1$, we have $\Zt{k}  \succeq \Yt{k} + 6 \gr{S}_{k+1,C_3} + 
2 \gr{J}_{k+1,C_3} \succeq 0$. 
\end{lem}

\noindent \textbf{Proof.}
First, we remark that
\begin{eqnarray}
\Omegat{k}(\xx^{-t}) & & = t(t+3) \xx^{-t-4} - t(2t+8) \xx^{-t-3} \cr
& & + t(t+8) \xx^{-t-2} - 7t \xx^{-t-1} + (5t-2k) \xx^{-t} - t \xx^{-t+1} \, .
\end{eqnarray}
Thus, using this (\ref{EQ:DELTA}) and (\ref{BB_K}),
  we have
\begin{eqnarray}
\Yt{k}(z) & & = \Big(6kb_k + 8(k+1) b_{k+1}\Big) \xx^{-3k-3} \cr
       & &  - \Big(15kb_k + 6(k+1)b_{k+1}\Big) \xx^{-3k-2} \cr
     & & + 21kb_k \xx^{-3k-1} -13kb_k \xx^{-3k} + 3kb_k \xx^{-3k+1} \, .
\label{\Yt{k}}
\end{eqnarray}
Similarly, we find
\begin{eqnarray}
\Zt{k}(z) & & = \Big(6kb_k + 8(k+1) b_{k+1}\Big) \xx^{-3k-3} \cr
     & & +\Big(2(4k+3) \cpt{k+1} + 2(3k-1)\cpt{k}%
     - 16(k+1)b_{k+1} - 12kb_k\Big) %
     \xx^{-3k-2} \cr
     & & + \Big(8(k+1) b_{k+1} + 6kb_k %
   - 2(3k+2)\cpt{k+1} - 5(3k-1) \cpt{k}\Big) %
     \xx^{-3k-1} \cr
     & & + 7(3k-1) \cpt{k} - (13k-5) \cpt{k} \xx + (3k-1) \cpt{k} \xx^2 \, .
\label{Zt_k}
\end{eqnarray}
Rearranging (\ref{\Yt{k}}), we obtain
\begin{eqnarray}
\Yt{k}(z) & & = 3kb_k \xx^{-3k-2} \big( 2\xx^{-1} -5 \big) \cr
       & & + 2(k+1)b_{k+1}  \xx^{-3k-3} \big( 4 \xx^{-1} -3 \big) \cr
       & & + kb_k \xx^{-3k} \big( 21\xx^{-1} -13 \big) + 3kb_k \xx^{-3k+1}
\end{eqnarray}
and so $\Yt{k} \succeq 0$.
By (\ref{EQ:LEMMA_BOUND_S3}) and (\ref{EQ:LEMMA_BOUND_J3}),
we have 
 $\gr{S}_{k+1,C_3} + \gr{J}_{k+1,C_3} \preceq \frac{2 k b_k}{\xx^{3k+2}}$
Hence, 
\begin{equation}
\Zt{k} - \Yt{k} -  6 \gr{S}_{k+1,C_3} - 2 \gr{J}_{k+1,C_3} 
  \succeq \Zt{k} - \Yt{k} - 12 kb_k \xx^{-3k-2}
\end{equation}
and
\begin{eqnarray}
& & \Zt{k} - \Yt{k} -  6 \gr{S}_{k+1,C_3} - 2 \gr{J}_{k+1,C_3} \, \, \, \, %
\succeq \cr
& &   \Big( 2(4k+3) \cpt{k+1} + 2(3k-1) \cpt{k} %
- 9 kb_k -10(k+1)b_{k+1} \Big) \xx^{-3k-2} \cr
& & + \Big( 8(k+1)b_{k+1} -15kb_k %
 - 2(3k+2) \cpt{k+1} - 5(3k-1) \cpt{k} \Big) \xx^{-3k-1} \cr
& & + \Big(7(3k-1)\cpt{k} + 13kb_k \Big) \xx^{-3k} %
-\Big((13k-5) \cpt{k} +3kb_k \Big) \xx^{-3k+1}  \cr
& & + (3k-1) \cpt{k} \xx^{-3k+2} \, .
\end{eqnarray}
Rewriting, we have
\begin{eqnarray}
& & \Zt{k} - \Yt{k} -  6 \gr{S}_{k+1,C_3} - 2 \gr{J}_{k+1,C_3} \, \, \, \, %
\succeq \cr
& &  \Big(2(4k+3)\cpt{k+1} + 2(3k-1)\cpt{k} %
- 9kb_k -10(k+1) b_{k+1} \Big) (\xx^{-1}-2)^2   \cr
& & + \Big(2(13k+10)\cpt{k+1} + 3(3k-1)\cpt{k} %
- 51kb_k -32(k+1)b_{k+1} \Big) (\xx^{-1}-2)  \cr
& & + \Big( (44k+28)\cpt{k+1} + 9(3k-1)\cpt{k} %
- 69kb_k -40(k+1)b_{k+1}   \Big) \cr
& & + \Big( (13k-5)\cpt{k} + 3kb_k \Big) (T - 1) \cr
& & + (3k-1)\cpt{k} \xx^{-2} 
\end{eqnarray}
and by lemma \ref{LEMMA0}, (\ref{CPRIME_0}) and  
(\ref{EQ:B_K}) after some calculations we find
$\Zt{k} - \Yt{k} -  6 \gr{S}_{k+1,C_3} - 2 \gr{J}_{k+1,C_3} \succeq 0$
\ENDPROOF

\begin{lem} 
\label{LEMMA_D}
For all $t \in \coeff{1,\, k-1}$, if
\begin{equation}
 \left( \gr{W}_{t,C_3} - \frac{b_t}{{\xx}^{3t}} %
+ \frac{ \cpt{t} }{{\xx}^{3t-1}} \right)   \succeq  0
\end{equation}
then
\begin{equation}
    \Delta_{k+1} \left[ \gr{W}_{k+1,C_3} - \frac{b_{k+1}}{{\xx}^{3k+3}}  %
+ \frac{ \cpt{k+1} }{{\xx}^{3k+2}} \right]   \succeq  %
      \Omegat{k} %
 \left[ \gr{W}_{k,C_3} - \frac{b_{k}}{{\xx}^{3k}} %
 + \frac{ \cpt{k} }{{\xx}^{3k-1}} \right] \, . 
\label{LEMMA9b_WRIGHT}
\end{equation}
\end{lem}

\noindent \textbf{Proof.} 
Using lemmas \ref{LEMMA_A}, \ref{LEMMA_C} and
(\ref{AVEC_OP}), we infer that
\begin{eqnarray}
& & \Delta_{k+1}{\EuScript{L}}_{k+1} - \Omegat{k} {\EuScript{L}}_{k}  = 
- 6 \gr{S}_{k+1,C_3} -  2 \gr{J}_{k+1,C_3} + %
\Lambdat{k} - Y_k + Z_k  \cr
& - & {\xx}^{-3k-2} T^2/{\xx}^2 %
(9\mathcal{B}_k - 6\Ckt{k}/{\xx})  %
\succeq  (Z_k - Y_k) \succeq 0 \, . 
\end{eqnarray}
\ENDPROOF

\begin{lem}  
\label{LEMMA_E}
Let $n_0 = n_0(k) = \frac{3}{2} + \sqrt{(2k + \frac{9}{4})}$.
 If $k \geq 2$ and $ 0 \leq n \leq n_0(k)$ then the coefficients 
of $\EuScript{L}_k$.
\end{lem}

\noindent \textbf{Proof.} If $n < n_0(k)$ then ${n \choose 2} < n+k$ and
\textit{a fortiori} there is no $(k+1)$-cyclic connected graphs.
Let $k \geq 2$ and $n  < n_0(k)$. Since $n! \coeff{z^n} \gr{W}_{k,\xi}(z)=0$,
we have to prove only that
\begin{equation}
n! \coeff{z^n} \Big( \frac{\cpt{k}}{{\xx}^{3k-1}} %
- \frac{b_k}{{\xx}^{3k}}  \Big) \geq 0 \, ,
\label{APROVER}
\end{equation}
because $n! \coeff{z^n} \frac{b_k}{{\xx}^{3k}} \geq 0$. As
$\cpt{k} \geq c_k$,  it suffices to show that
\[
n! \coeff{z^n} \Big( \frac{c_k}{{\xx}^{3k-1}} %
- \frac{b_k}{{\xx}^{3k}}  \Big) \geq 0 \, .
\]
Let 
\begin{equation}
M(z) = 1+\sum_{n\geq 1} n^n \frac{z^n}{n!} \, ,
\label{M&Ms}
\end{equation} 
i.e., $M = \frac{1}{\xx} = \frac{1}{1-T}$.
Note that
if $n  < j$ and $t < n_0$ then $t < 3k-1$ and
lemma \ref{LEMMA_A} tells us that $(3k-t)c_k \geq 3kb_k$ and
${{3k-1} \choose {t}}c_k \geq {{3k} \choose {t}}b_k$. Thus,
\begin{eqnarray}
& & n! \coeff{z^n} \left[ \frac{c_k}{{\xx}^{3k-1}} %
- \frac{b_k}{{\xx}^{3k}}  \right]   =  \cr
& & \,\,\,\,\,\,\,\,\,\,\,\,  n! \coeff{z^n}  \left[ c_k(1+M(z))^{3k-1} %
- b_k(1+M(z))^{3k}  \right] =  \cr
& &  \,\,\,\,\,\,\,\,\,\,\,\, \sum_{t=0}^{n_0} n! \coeff{z^n} 
\left[{{3k-1} \choose {t}}c_k - {{3k} \choose {t}}b_k \right] M(z)^t \, .
\end{eqnarray}
\ENDPROOF

We are now ready to prove (\ref{COEFF_INEGALITES}).

\begin{lem} 
\label{LEMMA_F}
For all  $k \geq 1$, the formal power series 
${\EuScript{L}}_k$ satisfies
\[
{\EuScript{L}}_k(z) = \gr{W}_{k,C_3}(z) - \frac{b_k}{{\xx}^{3k}} + %
\frac{\cpt{k}}{{\xx}^{3k-1}} \succeq 0\, .
\]
\end{lem}

\noindent \textbf{Proof.} 
First, $\EuScript{L}_1 \succeq 0$ by (\ref{BICYCLIC-MC3FREE}).
 Suppose that $\EuScript{L}_i \succeq 0$ 
for all $i \in \coeff{1,k-1}$ and  we have to show 
that $\EuScript{L}_k \succeq 0$.
Hence,  we can use lemma \ref{LEMMA_D}.
By definition,
\[
\Omegat{k-1} \EuScript{L}_{k-1}(z)  = %
\big( \V_z^2 - 3 \V_z - 2(k-1) \big) \EuScript{L}_{k-1}(z) +%
 2 \big( \V_z W_{0,C_3}(z) \big) %
\big( \V_z  \, \EuScript{L}_{k-1}(z) \big) \, .
\]
If $n > n_0(k)$ then $n^2 -3n -2k > 0$ and we have
\[
\begin{array}{ccc}
\coeff{z^n} \Omegat{k-1} \EuScript{L}_{k-1}(z) &  = & %
(n^2 -3n -2k+2) \coeff{z^n} \EuScript{L}_{k-1}(z) \cr
 & + & 2 \coeff{z^n} %
\big( \V_z W_{0,C_3}(z) \big) %
\big( \V_z  \, \EuScript{L}_{k-1}(z) \big) \geq 0
\end{array}
\]
Lemma \ref{LEMMA_D} tells us that ${\Delta_{k} \EuScript{L}_{k}} \geq
0$. Taking into account the definition of $\Delta$ given by
(\ref{OP:DELTA_K}), we obtain for  $n \geq n_0(k-1)$~:
\begin{equation}
2 (n+k) \coeff{z^n} \EuScript{L}_{k} %
\geq 2\coeff{z^n}{T \V_z\EuScript{L}_k}= %
2 \sum_{s=1}^{n-1} {n \choose s} s(n-s)^{n-s-1} %
\coeff{z^s} \EuScript{L}_k(z) \, .
\label{433333}
\end{equation}
And lemma \ref{LEMMA_E} leads to $\coeff{z^n} \EuScript{L}_{k}(z) \geq 0$,
if $n < n_0(k)$. Since $n_0(k-1) < n_0(k)$ we can infer by induction 
on $n$ using (\ref{433333}) that $\EuScript{L}_k \succeq 0$. \ENDPROOF

\subsection{Asymptotic results}
\label{SEC:N_UN_TIERS}
Denote by $c(n,n+k)$ the number of connected graphs having
$n$ vertices and $n+k$ edges.
Our aim of this paragraph is to 
establish that the number $c_\xi(n,n+k)$ of $\xi$-free connected graphs
with $n$ vertices and $n+k$ edges is asymptotically
the same as $c(n,n+k)$ whenever $k=o(n^{1/3})$.
Combining lemmas \ref{TREE_POLYNOMIAL_INFINITY},
 \ref{LEMMA_A} and \ref{LEMMA_F}, we obtain
the following important results:
\begin{thm} \label{THEOREM_ASYMPT1}
Almost all graphs having $n$ vertices and
$n+k$ edges are triangle-free when $n, \, k \rightarrow \infty$
but $k=o(n^{1/3})$.
\end{thm}

\noindent \textbf{Proof.} On one hand, lemma \ref{TREE_POLYNOMIAL_INFINITY}
shows that if $a \equiv a(n) \rightarrow 0$ as $n \rightarrow \infty$,
and if $b_1$ and $b_2$ are two fixed numbers such that $b_1 < b_2$,
then we have $t_n(an+\beta_1) \ll t_n(an+\beta_2)$ since
in  (\ref{EQ:TREE_POLYNOMIAL_INFINITY}) we obtain a factor
$(1-u_0)^{(1-\beta)} = %
(\sqrt{a(1+\frac{a}{4})} - \frac{a}{2})^{(1-\beta)} = %
a^{\frac{1-\beta}{2}} + O(a)$.
On the other hand, we have
\[
kb_k \leq \cpt{k} \leq \frac{25}{5}kb_k 
\]
and
\[
\frac{b_k}{{\xx}^{3k}} - \frac{\cpt{k}}{{\xx}^{3k-1}} \preceq %
\widehat{W}_{k,C_3} \preceq \frac{b_k}{{\xx}^{3k}}
 \, \, (k \geq 1) \, .
\]
Since $\cpt{k} = c_k + O((k-1)b_{k-1})= O(kb_k)$,
we have to find the values of $k$ for which
\[ k b_k t_n(3k-1) \ll b_k t_n(3k) \, . \]
We will use formula (\ref{EQ:TREE_POLYNOMIAL_INFINITY})
of lemma \ref{TREE_POLYNOMIAL_INFINITY}
to estimate $t_n(a\, n+\beta_1)$ and $t_n(a\, n+\beta_2)$,
with $an=3k, \, \beta_1=-1$, resp. $\beta_2= 0$. 
It proves convenient to compute
$\frac{kt_n(a\, n+\beta_1)}{t_n(a\, n+\beta_2)}$ and we have
\begin{eqnarray}
\frac{kt_n(a\, n+\beta_1)}{t_n(a\, n+\beta_2)} %
& = & \frac{k t_n(3k-1)}{t_n(3k)} \cr
          & = & k(1-u_0) = k(\sqrt{a}+O(a)) \cr
          & = & \frac{n}{3}(a^{\frac{3}{2}} + O(a^2)) \, .
\end{eqnarray}
Consequently, if $k=o(n^{1/3})$ the number 
$c_\xi(n,n+k)$ is asymptotically the same
as $c(n,n+k)$.
\ENDPROOF

Also, we have

\begin{thm}[Wright 1980] \label{WRIGHT_THM} 
As $n, \, k \rightarrow \infty$ but $k=o(n^{1/3})$, we have
\begin{eqnarray}
c(n,n+k) & = & d_k\, (3 %
\pi)^{1/2}(e/12k)^{k/2}n^{n+1/2(3k-1)} \cr
 & & \, \, \, \,  \, \, \, \, %
\times  \Big(1+O(k^{-1})+O(k^{3/2}/n^{1/2})\Big) \, 
\label{EQ:ASYMPT_N13} 
\end{eqnarray}
where $d_k = \frac{1}{2 \pi} + O(1/k)$.
\end{thm}
Note that  the value 
$d=\frac{1}{2 \pi}=\lim_{k\rightarrow \infty} \, d_k$
was independently found by Voblyi \cite{Vo87} and by Meertens \cite{BCM90}.

\noindent
As a corollary of theorems \ref{THEOREM_ASYMPT1} and \ref{WRIGHT_THM}, 
we obtain
\begin{cor} \label{THEOREM_ASYMPT2}
If $n, \, k \rightarrow \infty$ but $k=o(n^{1/3})$
the asymptotic number of \\ 
$(n,n+k)$ triangle-free connected graphs
is given by 
\begin{equation}
d_k\, (3 %
\pi)^{1/2}(e/12k)^{k/2}n^{n+1/2(3k-1)} %
  \Big(1+O(k^{-1})+O(k^{3/2}/n^{1/2})\Big) \, .
\end{equation}
\end{cor}

\section{Random graphs and forbidden subgraphs}
\label{SEC:RANDOM-GRAPHS}
As shown in \cite{FKP89,JKLP93},
the machinery of generating functions permits to study the limit
distribution of random graphs and multigraphs with great precision.
In this section, we will show that probabilistic results on random \
$\xi$-free
graphs and multigraphs can be obtained when looking at the form of their
generating functions, mainly looking at the so-called \textit{leading
coefficients} of their decompositions into tree polynomials, i.e.,
using the results of the previous sections 
 and some analytical facts contained in \cite{JKLP93}.

We consider here two models of random graphs, namely the
\textit{permutation model} and the
\textit{multigraph process}. The idea is to
start with $n$ totally disconnected vertices and to add 
successive edges one at time and at random \cite{ER59,ER60}. 
In the first model, also called \textit{graph process}, we consider
all $N={n \choose 2}$ possible edges $x \rbar y$ with $x <y$
which are introduced in random order, allowing all $N!$
permutations with the same probability. 

In the second model, also called \textit{uniform model},
ordered pairs $\langle x,y \rangle$ are generated
repeatedly ($1 \leq x, y \leq n$) and the edge $x \rbar y$ is added
to the multigraph. Thus, this process can generate
self-loops and multiple edges. Remark that we follow
Janson \textit{et al.} and for purposes of analysis, 
we assign a \textit{compensation factor} to a multigraph $M$, viz.
a multigraph $M$ on $n$ labelled vertices can be defined by a symmetric
$n\times n$ matrix of nonnegative integers 
$m_{xy}$, where $m_{xy}=m_{yx}$ is the number of undirected edges
$x\rbar y$ in~$G$. The {\it compensation factor\/} associated to $M$
is given by
\begin{equation}
\kappa(M)=
 1\left/\prod_{x=1}^n\left(2^{m_{xx}}\prod_{y=x}^nm_{xy}!\right)\right.
\label{COMPENSATION_FACTOR}   
\end{equation}
Thus, if  $m=\sum_{x=1}^n\sum_{y=x}^nm_{xy}$ is the total number of edges,
the number of sequences $\langle x_1,y_1\rangle\langle x_2,y_2\rangle
\,\ldots\,\langle x_m,y_m\rangle$ that lead to~$M$ is then exactly
\begin{equation}
2^m\,m!\,\kappa(M) \,.
\label{COMPENSATION_FACTOR2}
\end{equation}
(We refer to \cite[Sect. 1]{JKLP93} for more details about $\kappa$.)

At generating function level, it follows that 
after adding $m$ edges, the uniform model
on $n$ vertices will produce a 
multigraph  in a family $\mathcal{F}$ with 
probability
\begin{equation}
{2^m\,m!\,n!\over n^{2m}}\,\,[w^mz^n]\,\,F(w,z)\,.
\label{PROBA-MULTIGRAPH}
\end{equation}
Similarly, if ${\mathcal{F}}$ is a family of
graphs with labelled vertices, the probability that $m$ steps 
of the permutation model will produce
a graph in $\mathcal{F}$ is
\begin{equation}
{n!\over{N\choose m}}\,[w^mz^n]\,F(w,z)\,,\qquad N={n\choose 2}\,.
\label{PROBA-GRAPH}
\end{equation}

In \cite[Theorem 5]{JKLP93}, the authors
proved that only leading coefficients of $t_{n}(3k)$ are 
relevant to compute
the probability that randomly generated graphs or multigraphs
will produce $r_1$ bicyclic components, $r_2$ 
tricyclic components, $\cdots$ 
%
%
We have the following results about $\xi$-free components and
random graphs:

\begin{thm} \label{THEOREM-RG-XI-FREE}
The probability that a random graph or 
multigraph with $n$ vertices and $\frac{n}{2}$ edges has
only acyclic, unicyclic, bicyclic components all triangle-free is
\begin{equation}
\sqrt{\frac{2}{3}}\cosh \left( \sqrt{\frac{5}{18}}\right) e^{-\frac{1}{6}%
}+O(n^{-1/3})\approx 0.789... \, .
\label{PROPOSITION1}
\end{equation}
More generally, let 
$\Theta = \{ p \in \mathbb{N}, p \geq 3 \mbox{ and } C_p \in \xi \}$.
The probability that a random graph or multigraph with $n$
vertices and $\frac{n}{2}$ edges has only acyclic, unicyclic,
bicyclic components all $C_{p}$-free, $p\in \Theta$, is
\begin{equation}
\sqrt{\frac{2}{3}}\cosh \left( \sqrt{\frac{5}{18}}\right) %
e^{-\sum_{p\in \Theta }\frac{1}{2p}}+O(n^{-1/3}) \, .
\label{PROPOSITION11}
\end{equation}
\end{thm}

\noindent \textbf{Proof.} This is a corollary of \cite[eq (11.7)]{JKLP93}
using the formulae (\ref{ACYCLIC-MC3FREE}), (\ref{ACYCLIC-GC3FREE})
and  (\ref{BICYCLIC-GC3FREE-POLYNOMIAL}).
Incidentally, random graphs and multigraphs have the same asymptotic
behavior as shown by the proof of \cite[Theorem 4]{JKLP93}. As 
multigraphs graphs without cycles of length $1$
 and $2$, the forbidden cycles of length $1$ and $2$ bring a
factor $e^{-3/4}$ which is cancelled by a factor $e^{+3/4}$ because of the
ratio between weighting functions that convert the EGF of graphs and
multigraphs into probabilities. Indeed, 
formulae (\ref{PROBA-MULTIGRAPH}) and (\ref{PROBA-GRAPH}) are 
asymptotically related by the formula 
\begin{equation}
{ {n \choose 2} \choose m} =\left( \frac{n^{2m}}{2^{m}m!}\right) %
\exp{\big(-\frac{m}{n}-\frac{%
m^{2}}{n^{2}}+O(\frac{m}{n^{2}})+O(\frac{m^{3}}{n^{4}})\big)},\, m \leq {n
\choose 2} \, .
\label{RAPPORT-GRAPH/MULTI}
\end{equation}

The situation changes radically when cycles of length greater to or less
than $3$ are forbidden. Equations (\ref{ACYCLIC-MC3FREE}), (\ref
{ACYCLIC-GC3FREE}) and the ``significant coefficient'' $\frac{5}{24}$ of $%
t_{n}(3)$ 
in (\ref{BICYCLIC-GC3FREE-POLYNOMIAL}) and the demonstration of \cite[Lemma 3]
{JKLP93} show us that the term $-\frac{T(z)^p}{2p}$, introduced in (\ref
{ACYCLIC-MC3FREE}) and (\ref{ACYCLIC-GC3FREE}) for each forbidden $p$-gon,
simply changes the result by a factor of $e^{-1/2p}+O(n^{-1/3})$.\ENDPROOF

The example of forbidden $p$-gon suggests itself for a
generalization.

\begin{thm}
\label{THEOREM4} \textit{Let} $\xi
=\{H_{1},H_{2},H_{3},...H_{q}\} $ \textit{be a finite collection of
multicyclic connected graphs or multigraphs}. \textit{Then the probability
that a random graph with }$n$ \textit{vertices and} $\frac{1}{2}n+O(n^{%
\frac{1}{3}})$ \textit{edges} \textit{has} $r_{1}$ \textit{bicyclic
components}, $r_{2}$ \textit{tricyclic components},$\cdots $, $(k+1)$-%
\textit{cyclic components}, \textit{all components }$%
\{H_{1},H_{2},H_{3},...H_{q}\}$\textit{-free} 
\textit{and no components of
higher cyclic order is}
\begin{equation}
\big(\frac{4}{3}\big)^{r}\exp{\big(-\sum_{p\in \Theta }\frac{1}{2p}\big)}\,%
\sqrt{\frac{2}{3}}\,\frac{b_{1}^{r_{1}}}{r_{1}!}%
\,\frac{b_{2}^{r_{2}}}{r_{2}!%
}\,\,\cdots \,\,\frac{b_{k}^{r_{k}}}{r_{k}!}\frac{r!}{(2r)!}+O(n^{-1/3})
\label{theorem1}
\end{equation}
\textit{where }$\Theta =\{p\geq 3$ , $ \exists i\in [1,q]$ 
\textit{such that} $H_{i}$ \textit{is a} $p$\textit{-gon}$\}$.
\end{thm}

Theorem \ref{THEOREM4} raised a natural question. Under what
conditions on the 
forbidden configurations of graphs will the coefficients $(b_{i})$ 
change? The theorem \ref{THEOREM5}  below shows 
that a sufficient condition to 
change a coefficient $b_{i}$ of (\ref{theorem1}) is
 that $\xi$ must contain  all graphs 
\textit{contractible} to a certain $i$-excess graph $H$.

\begin{thm} \label{THEOREM5} Let  $H$ be a $k$-excess 
multicyclic graph (resp. multigraph) with $k>0$. Suppose
 that $c(H)\, n!$ is the number of ways to label $H$ 
(for example $c(K_{4})=1/24$). 
Denote by $\mathcal{A}_{k}^{(H)}$ the set of all $k$-excess 
graphs contractible to $H$. Then the probability
that a random graph (resp. multigraph) with $n$ vertices and 
$m(n)=\frac{n}{2}+O(n^{1/3})$ edges has $r_{1}$ bicyclic, $r_{2}$ 
tricyclic, ..., $r_{p}$ $(p+1)$-cyclic components, all
without component isomorphic to any member of the set 
$\mathcal{A}_{k}^{(H)}$ and with $r=r_1+2 r_2+\cdots+p r_p$ is 
\begin{equation}
\big(\frac{4}{3}\big)^{r}\sqrt{\frac{2}{3}}\,\frac{b_{1}^{r_{1}}}{r_{1}!}\,%
\cdots \,\,\frac{b_{k-1}^{r_{k-1}}}{r_{k-1}!}\,%
\frac{(b_{k}-c(H))^{r_{k}}}{r_{k}!}\,\frac{b_{k+1}^{r_{k+1}}}{r_{k+1}!}%
\,\,\cdots \,\,\frac{b_{p}^{r_{p}}}{r_{p}!}\frac{r!}{(2r)!}+O(n^{-1/3}) \, .
\end{equation}
\end{thm}
\noindent \textbf{Proof.} The EGF associated to $\mathcal{A}_{k}^{(H)}$ 
is simply 
\begin{equation}
A_{k}^{(H)}(w,z)=w^k \, c(H) \, \frac{T(wz)^n}{(1-T(wz))^{3k}} \, .
\label{proof of THEOREM5}
\end{equation}
Thus in (\ref{theorem1}) if we want to avoid all graphs \textit{%
contractible} to $H$, we have to subtract (\ref{proof of THEOREM5}) from
 the EGF of connected $k$-excess graphs.  $\ENDPROOF $

Note that in \cite[lemma 3]{JKLP93}, theorems
\ref{THEOREM-RG-XI-FREE}, \ref{THEOREM4}  and  \ref{THEOREM5},
the number of edges $m=m(n)$ varies from $\frac{n}{2}$
to  $\frac{n}{2}+ O(n^{1/3})$. The discrepancy in the windows is 
a consequence of the parameter $\mu$ in \cite[lemma 3]{JKLP93},
where $m(n) = \frac{1}{2}n(1+\mu n^{-1/3})$ and   $|\mu| \leq n^{1/12}$.
Hence, when choosing very small $\mu$, such as $\mu=O(n^{-1/3})$, one
can get results like theorems 4-5 in \cite{JKLP93} or
theorems \ref{THEOREM-RG-XI-FREE}, \ref{THEOREM4} and \ref{THEOREM5}
here.

\begin{ack}
The authors wish to thank C. Banderier, P. Flajolet  and
G. Schaeffer for helpful discussions and encouragements 
relating to this research and also all the anonymous referees for
their efforts reading and improving the quality of this paper.
\end{ack}

\bibliographystyle{alpha}
\bibliography{mabib}

\end{document}